\documentclass[a4paper,showkeys,nofootinbib]{revtex4}
\pdfoutput=1
\usepackage[utf8]{inputenc}
\usepackage{graphicx,bm,color}
\usepackage{amsmath}
\usepackage{slashed,verbatim}
\usepackage{subfigure}
\usepackage{hyperref}
\usepackage{gensymb}
\usepackage{footmisc}

\begin{document}
                                                                                                                                                                                                                     
\title{Analysis of \textit{Fermi}-LAT data from Tucana-II: Possible constraints on the Dark Matter models with an intriguing hint of a signal}

\author {Pooja Bhattacharjee$^{1,2}$}
\email []{pooja.bhattacharjee@jcbose.ac.in}
\author {Pratik Majumdar$^{3}$} 
\email []{pratik.majumdar@saha.ac.in}
\author {Sayan Biswas$^{4}$}
\email []{sayan@rri.res.in}
\author {Partha S.~Joarder$^{1,2}$}
\email []{partha@jcbose.ac.in} 
\affiliation{ $^{1}$ Department of Physics, Bose Institute, 93/1 A.P.C. Road, Kolkata 700009, India} 
\affiliation{ $^{2}$ Centre for Astroparticle Physics and Space Science, Bose Institute, Block EN, Sector V, Salt Lake, Kolkata 700091, India}
\affiliation{ $^{3}$ Saha Institute of Nuclear Physics, HBNI, 1/AF, Bidhannagar, Kolkata 700064, India}
\affiliation{ $^{4}$ Raman Research Institute, C.V. Raman Avenue, Sadashivanagar, Bangalore 560080, India}

\begin{abstract}
\noindent Tucana-II (Tuc-II), a recently discovered and confirmed Ultra Faint Dwarf Spheroidal galaxy, has a high mass to light ratio as well as a large line-of-sight stellar velocity dispersion, thus making it an ideal candidate for an indirect dark matter (DM) search. In this paper, we have analyzed nine years of $\gamma$-ray data obtained from the \textit{Fermi}-LAT instrument from the direction of Tuc-II. The fact that a very weak significant $\gamma$-ray excess ($2.2\sigma$) over the background of Tuc-II have been detected from the location of this galaxy. We have observed that this excess of $\gamma$-ray emission from the of location Tuc-II rises with longer periods of data. If WIMP pair annihilation is assumed for this faint emission, for $b\bar{b}$ annihilation channel the test statistics (TS) value peaks at DM mass $\sim$ 14 GeV and for $\tau^{+}\tau^{-}$ annihilation channel it peaks at DM mass 4 GeV. It is then called for an estimation of the $95\%$ confidence level upper limit of the possible velocity weighted self-annihilation cross-section of the DM particles (WIMPs) within Tuc-II by fitting the observed $\gamma$-ray flux with spectra expected for DM annihilation. The estimated upper limits of the cross-sections from Tuc-II are then compared with two other dwarf galaxies that are considered to be good DM candidates in several studies. We have also compared our results with the cross-sections obtained in various popular theoretical models of the WIMPs to find that our results impose reasonable tight constraints on the parameter spaces of those DM models. In the concluding section, we compared our results with the similar results obtained from a combined dSph analysis by the \textit{Fermi}-LAT collaboration as well as the results obtained from the studies of DM in the dwarf galaxies by the major ground-based Cherenkov experiments.

\end{abstract}
\maketitle

\section{Introduction}

\noindent The astrophysical and cosmological observations (e.g.~\cite{bib:kom, bib:ade}) strongly suggest that some kind of non-luminous and non-baryonic matter, namely the Dark Matter (DM), constitutes almost $75\%$ of the total matter density of the Universe. Regarding the physical nature of such DM, cosmological N-body simulations (e.g.~\cite{bib:die08, bib:spring08}) usually, favor a cold DM (CDM) scenario to explain the formation of the large scale structure of the Universe. In addition, the extension of the Standard Model (SM) of particle physics predicts that CDM could possibly consist of some form of massive, non-baryonic and neutrally charged particles, namely weakly interacting massive particles (WIMPs). WIMPs self pair annihilation (with their masses lying in the range of a few tens of GeV to a few hundreds of TeV) explains the thermal relic abundances (i.e. $<\sigma~v> \approx 3\times 10^{-26}~{\rm {cm}}^{3}~{\rm s}^{-1}$) and are consistent with the currently observed mass density of CDM; e.g..~\cite{bib:abdo, bib:ste, bib:jung}. Such pair-annihilation of the WIMPs may be one of the excellent source of indirect dark matter search~\cite{bib:abdo}. \\

\noindent DM pair-annihilation rate and the resulting flux of $\gamma$-photons are likely to be proportional to the square of the DM density. The dwarf Spheroidal satellite galaxies (dSph) of the Milky Way (MW) that are the densest DM regions in the galactic halo \cite{bib:alan}, usually lie away from the direction of the central region of the galaxy. Those dSphs are not too far ($\sim (20-200)$~kpc) from the earth \cite{bib:paitek} and they have low content of gas and dust \cite{bib:Grebel}, therefore, making them the potentially excellent targets for an indirect search of DM through the detection of the high energy $\gamma$-rays arising from the WIMP annihilations \cite{bib:abdo, bib:evansarkar, bib:Bonnivard}. Over about a decade from now, the Sloan Digital Sky Survey (SDSS) \cite{bib:york, bib:bel10}, the Panoramic Survey Telescope and Rapid Response System (Pan-STARRS) \cite{bib:kai, bib:lae1, bib:lae2}, the Dark Energy Survey (DES) \cite{bib:abb, bib:bec, bib:kop, bib:kim1} experiment and certain other surveys by using the Dark Energy Camera at Cerro Tololo~\cite{bib:kim2, bib:mar2} have discovered a new class of dSphs, namely the Ultra Faint Dwarf galaxies (UFDs) that have extremely low stellar contents and densities. The UFDs are dominated by old ($\gtrsim 12$~Gyr) stellar populations with large velocity dispersions that possibly indicate the existence of substantial DM components in those UFDs \cite{bib:koch}. The recent N-body simulations also indicate the existence of a huge number of DM sub-halos around the MW’s halo\cite{bib:drlicathesis, bib:kuhlennew} and amongst them, few hundreds of these sub-halos might be massive enough to host a dwarf galaxy \cite{bib:kuhlennew}. Thus, studying these old and metal-poor UFDs provide us a deep understanding on the nature of the ancient galaxies \cite{bib:sim, bib:kirby13} that were accreted to form the Milky Way halo \cite{bib:freb, bib:belokurovnew} and the origin of the chemical abundances of the stellar population of Milky Way halo\cite{bib:stark}. Hence, with inferred mass-to-light ratios reaching up to $\sim 3000~M_{\odot}/L_{\odot}$, the UFDs are, therefore, considered to be the best tracers of early DM sub-halos in the Universe as  predicted by the $\Lambda$CDM cosmological models~\cite{bib:kim17, bib:kuhlennew, bib:belokurovnew, bib:drlicathesis}. Recently, a joint DES-\textit{Fermi} collaboration~\cite{bib:albert17} has examined the $\gamma$-ray signatures of the WIMP pair-annihilations from about $45$ UFDs. The aim of their study was to re-examine the constraints imposed on various theoretical WIMP models by an earlier analysis \cite{bib:ack, bib:acknew, bib:acknew1, bib:abdo} of the $\gamma$-ray data from $15$ confirmed dSphs performed by the \textit{Fermi} collaboration. At present, two new sky survey programs, namely the Large Synoptic Survey Telescope (LSST) and the Wide-Field Infrared Survey Telescope, are being undertaken/planned to search for more UFD/dSph candidates in the galactic halo and in its neighborhood for a comprehensive study of the DM in the Universe~\cite{bib:chen}. \\

\noindent Motivated by such increasing interest in the indirect search for DM in the UFDs/dSphs, in the present paper, we focus our attention to a recently discovered dwarf satellite, namely Tucana-II (Tuc-II; DES~J2251.2-5836) \cite{bib:kop, bib:bec, bib:dril}. Tuc-II has already been confirmed to be a UFD (and not a part of any globular cluster) in Ref.~\cite{bib:wal}, principally because of its large projected half light radius, the large velocity dispersion of its member stars, its luminosity-metallicity relation and also because of its large dynamical mass to light ratio, all of which conform to the well-established values of the dwarf galaxies~\cite{bib:gilmore07, bib:kirby13, bib:mateo, bib:wal, bib:Ji}. Tuc-II may, as well, be a member of the Magellanic group as it is only about ~19 kpc away from the Large Magellanic Cloud (LMC) and about ~37 kpc away from the Small Magellanic Cloud (SMC)~\cite{bib:bec}. The outer region of Tuc-II appears to be in an elongated and distorted shape but the observational noise could, as well, be the reason for such a distortion \cite{bib:bec, bib:mar3, bib:mun}. Considering various observationally inferred parameter values of Tuc~II, Walker \textit{et al.}, 2016 \cite{bib:wal} had suggested that this UFD may not exhibit one of the strongest DM annihilation signal, but may contribute meaningfully to the analysis of stacked data from multiple sources including Tuc-II. Many groups have already studied the Tuc-II in search of dark matter self-annihilation \cite{bib:drlicanew, bib:hoopernew, bib:albert17, bib:calorenew}. \\

\noindent In this paper, we performed a data analysis of Tuc~II with 9 years of \textit{Fermi}-LAT data to search the dark matter signal. We observed a faint emission from the direction of Tuc-II for two WIMP pair annihilation channels and we showed the significant increment of the test statistic (TS) peak values from the direction of Tuc-II with larger periods of data. Both the test statistic (TS) and the statistical significance (i.e., the $p$-value) of the best-fitted spectra for this excess emission, apparently favors a DM annihilation spectra over a simple power-law spectra, that seems to undermine an astrophysical origin of the emission. Admittedly, the significance of the excess emission in the direction of Tuc-II, obtained in this paper, is much weaker than the threshold of the $Fermi$-LAT's detection but it gives a hint of dark matter signal. Such emission from Tuc-II location has not been reported before by any other groups. Moreover, we calculated the possible upper limits of the velocity-averaged pair-annihilation cross-section $<\sigma v>$ of the WIMPs from Tuc-II and compared with the ones obtained from the analysis of Ursa Minor (UMi). Later, we also compared our results with the ones obtained from the recent DES-discovered UFD, namely Reticulum-II (Ret-II)~\cite{bib:kop, bib:kop2, bib:bec, bib:simonnew, bib:walkernew1}. The \textit{Fermi}-LAT data analysis of Ret-II has already been done in Refs.~\cite{bib:drlicanew, bib:hoopernew, bib:geringernew2, bib:albert17, bib:linew, bib:zhaonew} and that seems to exhibit a small excess of $\gamma$-ray signal of some significance over the background of Ret-II, thus making Ret-II an attractive source to search for the annihilation signals of DM. Our comparison, presented in this paper, showed that the constraints imposed by Tuc-II on the popular WIMP models for $b\bar{b}$ annihilation channel are more stringent than the ones expected in the case of Ret-II. 
Furthermore, with nine years of \textit{Fermi}-LAT data, we have compared the resulting LAT sensitivity for Tuc-II with predictions obtained from four theoretical models. For the sake of such study, we here assume a perfect spherical symmetry for Tuc-II and further assume Tuc-II to be in dynamic equilibrium with a negligible contribution to its significantly large, observed line-of-sight stellar velocity dispersion from the possible binary stellar motions in Tuc-II, thus assuming that the gravitational potential of Tuc-II is entirely dominated by DM~\cite{bib:wal}. \\

\noindent The paper is organized along the following line. After stating the observed properties of Tuc-II (in subsection 2.1) that are relevant for our study, we briefly describe the procedure for the analysis of \textit{Fermi}-LAT data from the direction of Tuc-II in subsection 2.2. In subsection 2.3, we estimate the upper limits of the $\gamma$-ray flux from Tuc-II by fitting Tuc-II with power-law spectral model with five different power indices. In subsection 3.1, we employ the NFW density profile to model the DM density in UFDs. In subsection 3.2, we first report a faint $\gamma$-ray emission ($2.2\sigma$) from Tuc-II possibly resulting from WIMP pair annihilation to $\tau^{+}\tau^{-}$ channel by using the DMfit Monte Carlo (MC) simulation package~\cite{bib:jel} and study the nature of this excess emission for three, six and nine years of Fermi-LAT data. Then we have studied the distribution of excess obtained from the location of Tuc-II and the possible reason responsible for such faint emission. Next we determine the possible upper limit of the $\gamma$-ray flux from Tuc-II. There, we also calculate the possible upper limits of the velocity-averaged pair-annihilation cross-section $<\sigma v>$ of the WIMPs for several important pair-annihilation channels and compare those cross-sections with the ones obtained from various theoretical WIMP models in section 3.2. Comparisons of the $<\sigma v>$ upper limit from Tuc-II, with the results obtained from Ret-II and UMi, are also presented in subsection 3.2 of this paper. Finally, a brief discussion of our results presented here \textit{vis-a-vis} the results obtained from UFDs/dSphs with the \textit{Fermi}-LAT and with various ground-based Cherenkov experiments at higher energies are presented in the concluding section 4.\\

\section{Analysis of Tuc-II}
\subsection{The relevant observed properties of Tuc-II}

\noindent A spectroscopic study of a number of stars in the direction of Tuc~II was undertaken by Walker \textit{et al.}, 2016 \cite{bib:wal} by the use of the Michigan Magellan Fibre System (M2FS). This study \cite{bib:wal}, along with the previous photometric results on Tuc-II~\cite{bib:kop, bib:dril}, could identify eight probable member stars of Tuc-II that were sufficiently well-resolved to determine an internal velocity dispersion but with large asymmetrical uncertainties, $\rm{\sigma_{v}}~=~8.6_{-2.7}^{+4.4}~{\rm {km}~{s}}^{-1}$ about a mean velocity of $-129.1_{-3.5}^{+3.5}~{\rm {km}~{s}}^{-1}$ in the solar rest frame. These and the other important physical properties of Tuc-II that have either been directly observed or have been inferred from the observations of Tuc-II by the authors of Refs.~\cite{bib:kop, bib:wal, bib:chitinew}, are tabulated in TABLE~1 for later reference.

\begin{table}[h!]
\begin{center}
\caption{Properties of Tuc-II}
\label{Table-1}
\begin{tabular}{|p{4 cm}|p{4 cm}|p{2 cm}|}
\hline
\hline
Property &  Value   & Reference  \\ 
\hline
Galactic longitude & $\rm{328.0863^{\circ}}$  & \cite{bib:kop} \\
\hline
Galactic latitude & $\rm{-52.3248^{\circ}}$ & \cite{bib:kop} \\
\hline
Heliocentric distance ([d]) & $\rm{57_{-5}^{+5}~\rm {kpc}}$ & \cite{bib:kop}\\
\hline
Metallicity ([$\rm{Fe/H}$]) & $\rm{<0.4}$ & \cite{bib:wal}\\ 
\hline
Projected half light radius ($\rm{R_{h}}$) & $\rm{165^{+27.8}_{-18.5}~pc}$ & \cite{bib:kop} \\
\hline
Maximum galactocentric angular distance in the sample of the observed member stars in Tuc-II, as measured from the observer's position ([$\theta_{\rm {max}}$]) & $0.30^{\circ}$ & \cite{bib:chitinew}\\ 
\hline
Square-root of the luminosity-weighted square of the line-of-sight stellar velocity dispersion ($\rm{\sigma_{v}}$) & $~\rm{8.6_{-2.7}^{+4.4}~km~s^{-1}}$ & \cite{bib:wal}\\
\hline
Mass within the projected half-light radius \Big($\rm{\frac{M_{1/2}}{M_{\odot}}}$\Big) & $\rm{~2.7_{-1.3}^{+3.1}~\times~10^{6}}$ & \cite{bib:wal}\\
\hline
Dynamical mass-to-light ratio \Big($\rm{(M/L_{v})_{1/2}}$\Big) & $\rm{~1913_{-950}^{+2234}~M_{\odot}~L_{\odot}^{-1}}$ & \cite{bib:wal}\\
\hline
\hline
\end{tabular}
\end{center}
\end{table}

\noindent In TABLE~1, $M_{\odot}$ and $L_{\odot}$ indicate the mass and the total luminosity of the Sun, respectively. Definitions of various other quantities displayed in TABLE~1 are given in \cite{bib:wal, bib:kop, bib:chitinew}; also see Refs.\cite{bib:walk09, bib:wolf, bib:evan1}.

\subsection{The \textit{Fermi}-LAT data analysis of Tuc-II}

\begin{table}
    \begin{center}
    \caption{Parameters used in \texttt{Science Tools} for \textit{Fermi}-LAT data analysis}
    \begin{tabular}{|llll|}
        \hline \hline
        {\bf Parameter for data extraction\footnote{\url{https://fermi.gsfc.nasa.gov/ssc/data/analysis/scitools/extract_latdata.html}}} & & &\\
        \hline\hline
        Parameter & Value & &\\
        \hline \hline
        Source & Tucana-II & &\\
        Right Ascension (RA) & 342.9796 & &\\
        Declination (DEC) & -58.5689 & &\\
        Radius of interest (ROI) &  $10^{\circ}$ & &\\
        TSTART (MET) & 239557418 (2008-08-04 15:43:37.000 UTC) & &\\
        TSTOP (MET) & 530362359 (2017-10-22 10:52:34.000 UTC) & &\\
        Energy Range & 100 MeV - 300 GeV  & &\\
        \textit{Fermi}-LAT Science Tool version & \texttt{v10r0p5}\footnote{\url{https://fermi.gsfc.nasa.gov/ssc/data/analysis/software/}} & &\\
        \hline \hline
        $~~~~~~~~~~~~~~~~~~~$\texttt{gtselect} for event selection\footnote{\url{https://fermi.gsfc.nasa.gov/ssc/data/analysis/scitools/help/gtselect.txt}\label{gtselect}}& & &\\
        \hline \hline
        Event class & Source type (128)\footnote{\url{https://fermi.gsfc.nasa.gov/ssc/data/analysis/documentation/Cicerone/Cicerone_Data_Exploration/Data_preparation.html}\label{datacut}} & & \\
        Event type & Front+Back (3)\footref{datacut} & &\\
        Maximum zenith angle cut & $90^{\circ}$\footref{datacut} & &\\
        \hline \hline
        $~~~~~~~~~~~~~~~~~~~$\texttt{gtmktime} for time selection\footnote{\url{https://fermi.gsfc.nasa.gov/ssc/data/analysis/scitools/help/gtmktime.txt}} & & & \\
        \hline \hline
        Filter applied & $\rm{(DATA\_QUAL>0)\&\&(LAT\_CONFIG==1)}\footnote{\url{https://fermi.gsfc.nasa.gov/ssc/data/analysis/scitools/data_preparation.html}\label{data_preparation}}$ & &\\
        ROI-based zenith angle cut & No \footref{data_preparation}  & &\\
        \hline \hline
        $~~~~~~~~~~~~~~~~~~~$\texttt{gtltcube} for livetime cube\footnote{\url{https://fermi.gsfc.nasa.gov/ssc/data/analysis/scitools/help/gtltcube.txt}} & & &\\
        \hline \hline
        Maximum zenith angle cut ($z_{cut}$) & $90^{\circ}\footnote{\url{https://fermi.gsfc.nasa.gov/ssc/data/analysis/documentation/Cicerone/Cicerone_Likelihood/Exposure.html}}$ & &\\
        Step size in $cos(\theta)$ & 0.025 & & \\
        Pixel size (degrees) & 1 & &\\
        \hline \hline
        $~~~~~~~~~~~~~~~~~~~$\texttt{gtbin} for 3-D (binned) counts map\footnote{\url{https://fermi.gsfc.nasa.gov/ssc/data/analysis/scitools/help/gtbin.txt}} & & & \\
        \hline \hline
        Size of the X $\&$ Y axis (pixels) & 140 & &\\
        Image scale (degrees/pixel) & 0.1 & &\\
        Coordinate system & Celestial (CEL) & &\\
        Projection method & AIT & &\\
        Number of logarithmically uniform energy bins & 24 & &\\ 
        \hline \hline
        $~~~~~~~~~~~~~~~~~~~$\texttt{gtexpcube2} for exposure map\footnote{\url{https://fermi.gsfc.nasa.gov/ssc/data/analysis/scitools/help/gtexpcube2.txt}} & & &\\
        \hline \hline
        Instrument Response Function (IRF) & $\rm{P8R2\_SOURCE\_V6}\footnote{\url{https://fermi.gsfc.nasa.gov/ssc/data/analysis/documentation/Pass8_usage.html}\label{pass8}}$ & & \\
        Size of the X $\&$ Y axis (pixels) & 400 & &\\
        Image scale (degrees/pixel) & 0.1 & &\\
        Coordinate system & Celestial (CEL) & &\\
        Projection method & AIT & &\\
        Number of logarithmically uniform energy bins & 24 & &\\ 
        \hline \hline
        $~~~~~~~~~~~~~~~~~~~$diffuse models and Source model XML file\footnote{\url{https://fermi.gsfc.nasa.gov/ssc/data/analysis/user/make3FGLxml.py}} & & &\\
        \hline \hline
        Galactic diffuse emission model & $\rm{gll\_iem\_v06.fits}\footnote{\url{https://fermi.gsfc.nasa.gov/ssc/data/access/lat/BackgroundModels.html}\label{background}}$ & & \\
        Extragalactic isotropic diffuse emission model & $\rm{iso\_P8R2\_SOURCE\_V6\_v06.txt}\footref{background}$ & &\\
        Source catalog & 3FGL & & \\
        Extra radius of interest &  $5^{\circ}$ & &\\
        Spectral model of Tucana-II & Power law (in Section-2.3) $\&$ DMFit Function (in Section-3.2) \footnote{\url{https://fermi.gsfc.nasa.gov/ssc/data/analysis/scitools/source_models.html}} & &\\
        \hline \hline
        $~~~~~~~~~~~~~~~~~~~$\texttt{gtlike} for likelihood analysis\footnote{\url{https://fermi.gsfc.nasa.gov/ssc/data/analysis/scitools/binned_likelihood_tutorial.html}} & & &\\
        \hline \hline
        Response functions & $\rm{P8R2\_SOURCE\_V6}\footref{pass8} $ & & \\
        Optimizer & NEWMINUIT & &\\
        \hline \hline

    \end{tabular}
    \end{center}
    \label{table_1}
\end{table} 
\noindent The \textit{Fermi}-LAT is a $\gamma$-ray space-based detector that scans the whole sky every 3 hour for an efficient study of the $\gamma$-ray sky in an energy range from about 20 MeV to 500 GeV. In this paper, we have used the recent version \texttt{v10r0p5} of the \texttt{Fermi ScienceTools} for the analysis of $\gamma$-ray data from Tuc-II. The above version allows us to use the pre-processed PASS 8 dataset of event class 128 that makes use of an improved instrument response function (IRF) P8R2\_SOURCE\_V6 of the LAT.\\

\noindent We have extracted nearly nine years (i.e. from 2008-05-04 to 2017-10-22) of LAT data in 100 MeV to 300 GeV reconstructed energy range within a $10^{\circ}~\times~10^{\circ}$ radius of interest (ROI) centered on the location of Tuc-II. In the source model, we have included Tuc-II along with all the point sources from 3FGL catalog within 15$^{\circ}$ of ROI from the position of Tuc-II. We have then performed the binned likelihood analysis on our extracted dataset with the `gtlike' tool~\cite{bib:cas, bib:matt} by following the instructions given in the \texttt{ScienceTools}. The spectral parameters of all the \textit{Fermi}-3FGL sources~\cite{bib:ace} within ROI, as well as, the normalization parameters of two diffuse models (i.e. $\rm{gll\_iem\_v06.fits}$ and  $\rm{iso\_P8R2\_SOURCE\_V6\_v06.txt}$) have been left free during the model fitting procedure. The rest of all the background sources within the $15^{\circ}~\times~15^{\circ}$ ROI have been kept fixed at their values given in the 3FGL catalog \cite{bib:ace}. All the required information related to our \textit{Fermi}-LAT analysis method is mentioned in TABLE~2.\\

\noindent In the following, we first model Tuc-II as a source having a power-law spectrum with each of the five different spectral indices and later in section 3.2, we go over to fit the $\gamma$-ray spectrum arising from the (assumed) DM-dominated Tuc-II with an MC-simulated DM self-annihilation spectrum generated by the use of the DMFit simulation tool-kit~\cite{bib:jel}.

\subsection{Results of the power-law modeling} 

\begin{figure}[h!]
\subfigure[]
 { \includegraphics[width=0.7\linewidth]{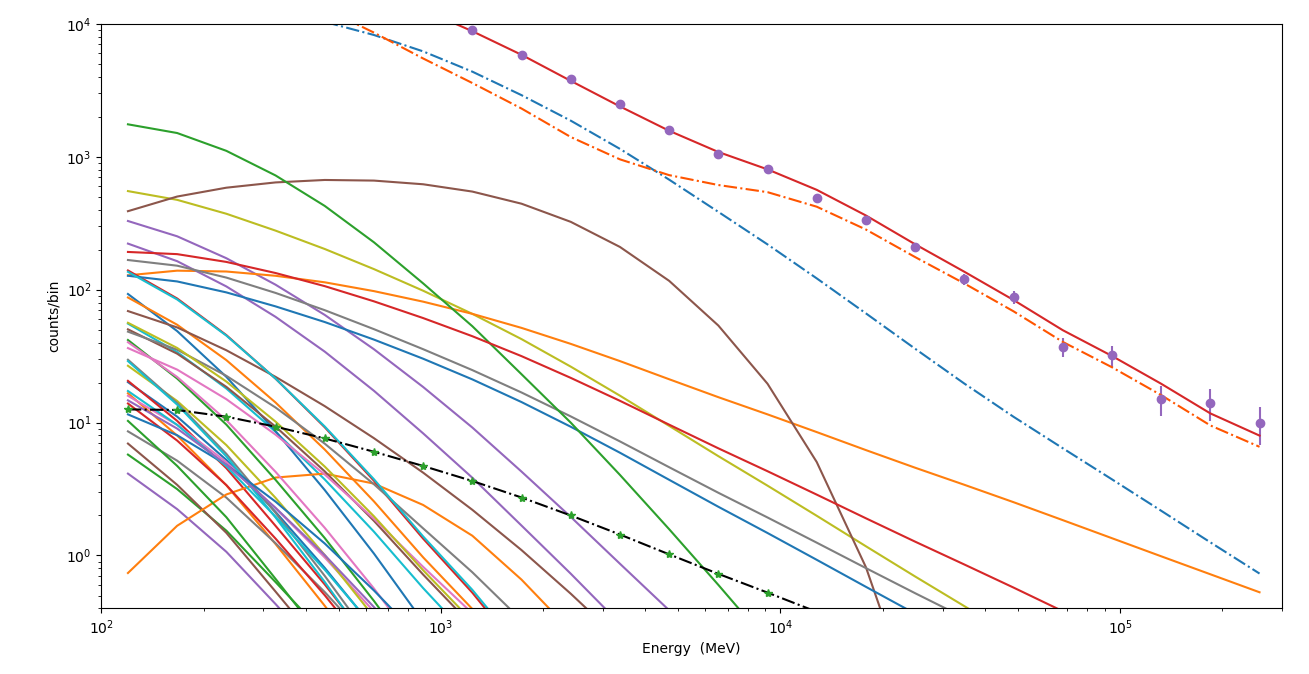}}
\subfigure[]
 { \includegraphics[width=0.8\linewidth]{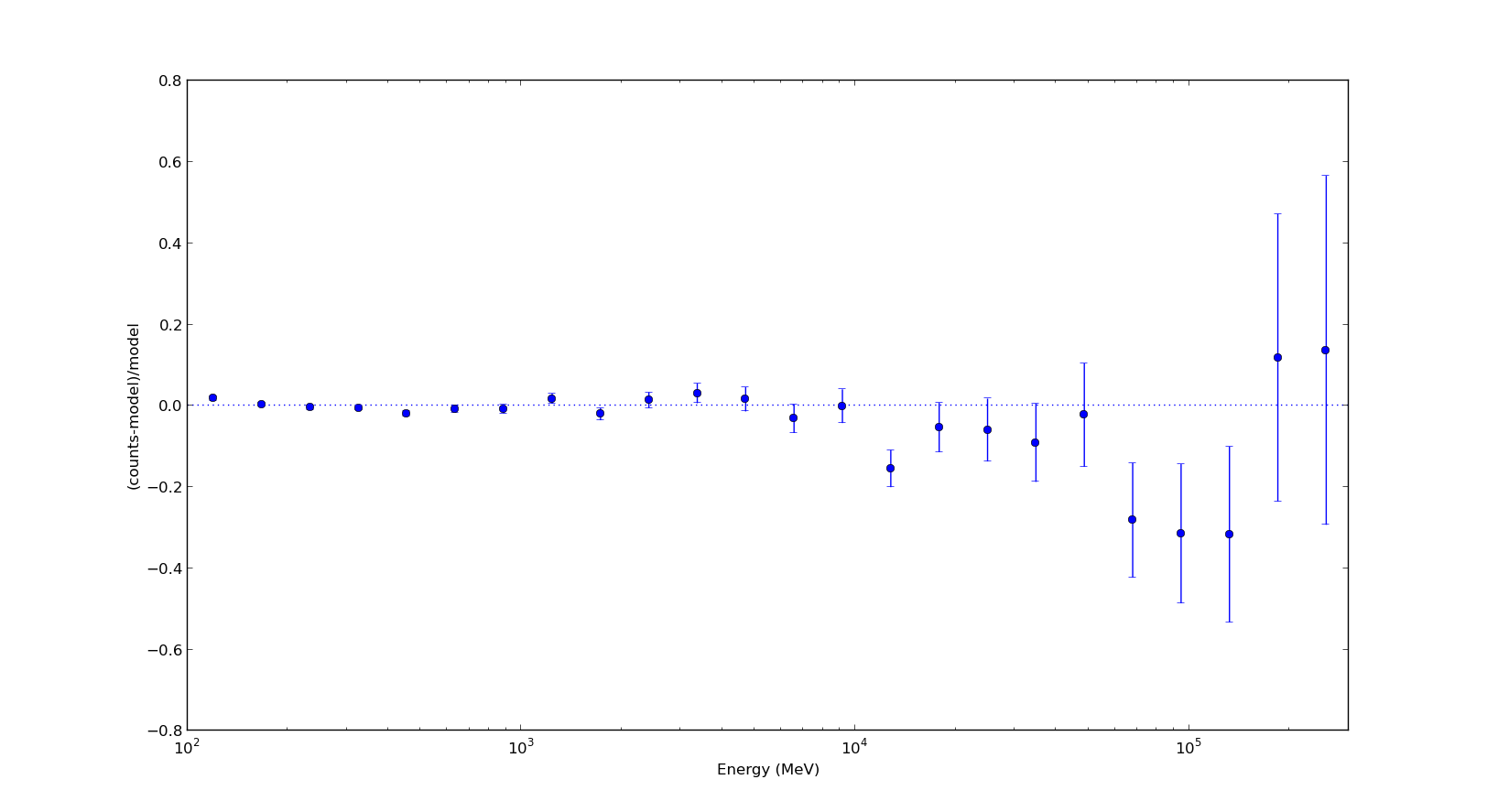}}
\caption{ Spectral fit to the counts per energy bin (Fig.~1(a)) and the corresponding residual plot (Fig.~1(b)) are displayed for all the sources within the chosen ROI centered on Tuc-II, in which the power-law source-spectral index of Tuc-II is taken as $\Gamma = 2$. In Fig.~1(a), the solid dark reddish-brown curve displays the best-fit total spectrum, along with the corresponding LAT-observed data points (in purple); the dot-dashed sky and orange curves display the galactic diffuse background and the isotropic background component, respectively; the dot-dashed black curve along with green points denotes the spectral fit of Tuc-II. The rest of the curves correspond to various point sources other than Tuc-II, lying within the ROI that are not distinctly labeled in Fig.~1(a).}
\end{figure}

\noindent The form of the differential photon flux for power-law modeling can be expressed as:
\begin{equation}
\frac{dN}{dA dE dt} = N_{0} \Big(\frac{E}{E_{0}}\Big)^{-\Gamma},
\end{equation} 

\noindent where, $dN$ denotes the number of photons lying between the energy range of $E$ and $E + dE$. $dA$ is the elemental area of detector, in which, photons are incident in time interval, $dt$.\\

\noindent From Eq.~(1) $N_{0}$ and $\Gamma$ define the normalization parameter and the spectral index of power-law modeling, respectively, while, $E$ denotes the reconstructed energy. For our analysis we have set the energy scale ($E_{0}$) at 100 MeV ~\cite{bib:abdo} and $dE$ is varied from 100 MeV to 300 GeV.\\

\noindent In Fig.~1(a), we display the spectral fit per energy bin of all the sources within the aforesaid ROI, along with the isotropic background component and the galactic diffuse background component. Similarly, Fig.~1(b) displays the corresponding residual plot for all the above sources in the given ROI. The horizontal axes in both these figures indicate the reconstructed energy $E$ of the $\gamma$-photons within the chosen range. While the above figures display only the results for modeling Tuc-II with a spectral index $\Gamma =2$ alone, we have actually repeated this spectral fitting procedure for each of the other values of the spectral index, namely $\Gamma = 1, 1.8, 2.2~{\rm {and}}~2.4$, as well.\\

\noindent The best-fitted values of the $N_{\rm {0}}$ and the TS obtained from Tuc-II for each of those spectral indices ($\Gamma$) are displayed in TABLE~3. The TS value is the ratio of the maximum likelihoods for two hypothesis, in which, $L_{\rm {(max, 1)}}$ denotes the maximum likelihood for full model and $L_{\rm {(max, 0)}}$ refers to the maximum likelihood for the null hypothesis. The expression of TS value is~$TS = -2\ln\Big(L_{\rm {(max, 0)}}/L_{\rm {(max, 1)}}\Big)$. Among those aforementioned spectral indices, $\Gamma = 1$ is preferred for its connection with the DM annihilation models in Ref.~\cite{bib:ess} and the other four $\Gamma$'s are chosen to examine the general astrophysical source spectrum. In TABLE~3, for $\Gamma = 1$, the error on $N_{\rm {0}}$ is slightly higher than the value $N_{\rm {0}}$ itself and it denotes the no-significance from the direction of Tuc-II and the TS values for other $\Gamma$s are also much less than the threshold-detection limit of \textit{Fermi}-LAT (i.e. TS$\ge$25).\\ 

\begin{table}[!h]
\begin{center}
\caption{Best-fitted normalization parameters ($N_{0}$) and the TS values obtained from the spectral fittings of the $\gamma$-ray flux from Tuc-II with five different spectral indices ($\Gamma$).}
\begin{tabular}{|p{3cm}|p{5cm}|p{3cm}|}
\hline 
\hline
Spectral~Index~($\Gamma$) & $\rm{N_{0} \times 10^{-5}~(cm^{-2}~s^{-1}~MeV^{-1})}$  & Test Statistic (TS) value \\
\hline 
$1$  & $(2.457\pm11.17)\times10^{-10}$ & 0.056   \\
\hline 
$1.8$  & $(1.173\pm1.126)\times10^{-7}$ & 1.215   \\
\hline 
$2$  & $(3.146\pm 2.565)\times10^{-7}$ & 2.077   \\
\hline 
$2.2$  & $(7.458\pm4.923)\times10^{-7}$ & 2.973   \\
\hline 
$2.4$ & $(1.433\pm0.839)\times10^{-6}$ & 3.592   \\
\hline 
\hline
\end{tabular}
\end{center}
\end{table}

\begin{table}[!h]
\caption{$95\%$ C.L. $\gamma$-ray flux upper limit obtained from the power-law spectral modelings of Tuc-II with five different spectral indices ($\Gamma$).}
\begin{center}
\begin{tabular}{|p{3cm}|p{8cm}|}
\hline 
\hline
Spectral~Index~($\Gamma$) & Flux~upper~limits~in~$\rm{95\%~C.L.~(cm^{-2}~s^{-1})}$ \\
\hline 
1 & $3.248\times10^{-11}$ \\
\hline 
1.8 & $4.484\times10^{-10}$ \\
\hline 
2 & $8.362\times10^{-10}$ \\
\hline 
2.2 & $1.401\times10^{-9}$ \\
\hline 
2.4 & $2.113\times10^{-9}$ \\
\hline
 \hline
\end{tabular}
\end{center}
\end{table}

\noindent Hence, we calculate the flux upper limit from Tuc-II over the entire reconstructed energy range ($0.1-300$)~GeV by the profile likelihood method \cite{bib:bar, bib:rol}. During the process of estimating the flux upper limits, all the normalization parameters along with two diffuse components are fitted continuously with the entire dataset until the logarithmic difference of two likelihood functions arrives at the value of $1.35$ \cite{bib:abdo} which corresponds to a one-sided $95\%$ C.L.\\

\noindent As no significant excess is observed at the location of Tuc-II, we have next estimated the 95$\%$ flux upper limits by using the semi-Bayesian method with flat prior. For very low data statistics system, the semi-Bayesian method is generally favored over likelihood profile \cite{bib:rol}. This  Bayesian method  is developed from Helene’s approach \cite{bib:helene} and is already implemented in the pyLikelihood module of \texttt{ScienceTools} as function bayesianUL() of python code `UpperLimits.py'. With this method, the flux upper limits in 95$\%$ C.L. is being estimated by integrating the whole likelihood profile, in which the integration was started from the lower bound of normalization parameter i.e. from $N_{0}=0$ without considering any specific distribution.\\

\noindent The 95$\%$ flux upper limits estimated from the semi-Bayesian method are displayed in TABLE~4 for each of the spectrum indices ($\Gamma$) considered above. In TABLE~4, we note that, the $95\%$ C.L. $\gamma$-flux upper limit for $\Gamma=1$ is almost 2 orders of magnitude lower than the one corresponding to $\Gamma=2.4$. This result is consistent with our previous work for Triangulum-II (Tri-II)~\cite{bib:poo}, in which we have also found that the flux upper limit in $95~\%~C.L.$ is increased by increasing the spectral indices. Here, we would like to add that though we have used the semi-Bayesian method for obtaining the flux upper limit, with profile likelihood method we also have obtained the same order of flux upper limits. They are hardly differed by 1.2 to 1.3 factor.\\

\noindent In section 3.2, we attempted to examine the dark matter signature, hence we have modeled Tuc-II with the $\gamma$-ray spectrum from DM annihilation (DMFit function) implemented in \textit{Fermi} \texttt{ScienceTools}. In that section, along with Tuc-II, we have introduced two other dwarf galaxies, namely, Ret-II and UMi and we have also followed the same analysis procedure for them that we have performed for Tuc-II (mentioned in TABLE~2).

\section{The $\gamma$-ray signature of the WIMP-annihilations in Tuc-II and the constraints on the DM models}

\subsection{Estimation of the flux of $\gamma$-rays from Tuc-II}
\noindent The expression for the differential photon flux, arising from WIMP pair-annihilations, in a DM source subtending a solid angle $\Delta \Omega$ at the observer's location is known (e.g.~\cite{bib:bal, bib:abdo}) to be
\begin{equation}
\phi_{\rm{WIMP}}(E, \Delta \Omega)~ = ~ \Phi^{pp}(E) \times J(\Delta \Omega),
\end{equation}
\noindent in which, $\Phi^{pp}(E)$, with the photon energy E, is the \textit{Particle physics factor}; whereas, $J(\Delta \Omega)$ in Eq.~(2) is the \textit{Astrophysical factor} or the J-factor. For estimating the $\gamma$-ray flux from Tuc-II, we have used the same approach as we used in our previous paper~\cite{bib:poo} but in the following sections, we reproduce brief descriptions of these factors for the sake of completeness. 

\subsubsection{\textbf{Particle physics factor}}

\noindent The expression for the particle physics factor that provides information regarding the properties of the initial and the final state particles in various possible WIMP pair-annihilation channels is given by~\cite{bib:abdo}

\begin{equation}
\Phi^{pp}(E)~ = ~ \frac{<\sigma v>}{8 \pi ~m^{2}_{\rm{WIMP}}} \sum_{f} \frac{dN_{f}}{dE}B_{f}.\\
\end{equation}

\noindent In Eq.~(3), $<\sigma v>$ denotes the thermally averaged product of the relative velocity between the WIMPs and their pair-annihilation cross-section \cite{bib:abdo}; whereas, $\frac{dN_{f}}{dE}$ and $B_{f}$ denotes the differential photon spectrum per DM annihilation and the branching ratio of a particular WIMP pair annihilation final state `$f$', respectively. 

\subsubsection{\textbf{Astrophysical factor (J-factor)}}

\noindent The expression for the J-factor in Eq.~(2) that contains the information regarding the astrophysical properties of the potential DM source (i.e., Tuc-II in the context of this paper), takes the form~\cite{bib:abdo}:
\begin{equation}
J(\Delta\Omega)=J(\lambda,\theta) = 2\pi {\int_{0}}^{\theta_{\rm {max}}} \sin{\theta} \int_{\lambda_{\rm{min}}}^{\lambda_{\rm{max}}} \rho^{2}(\sqrt{\lambda^{2} + d^{2} - 2\lambda d\cos{\theta}})d\lambda d\theta,
\end{equation}
\noindent Where, $\rho(r)$ is the radial distribution of the DM mass-density in UFDs. In Eq.~(4), $\lambda$ is the line-of-sight (l.o.s) distance, $d$ is the heliocentric distance and $\theta$ is the angle between the l.o.s and the center of UFDs, respectively. \\

\noindent We have used the simple analytic formula to calculate the J-factors provided by Evans \textit{et al.}, 2016; ref.~\cite{bib:evan1}. The formula is derived for the spherical Navarro-Frenk-White (NFW) model that follows the empirical relationship between enclosed mass, velocity dispersion and half-light radius. \\

\noindent The expression of NFW density profile is:
\begin{eqnarray}
\rho(r) &=& \frac{\rho_{0}r_{s}^{3}}{r(r_{s} + r)^{2}}
\end{eqnarray}

\noindent Where, $\rho_{0}$ and $r_{s}$ are the characteristic density and scale radius respectively and $r$ is the distance from the center of the galaxy. \\

\noindent The formula of J-factor from Evans \textit{et al.}, 2016 \cite{bib:evan1} is given as:
\begin{equation}
J= \frac{\pi \rho_{0}^{2} r_{s}^{3}}{3 d^{2} \Delta^{4}} \Big[2y(7y-4y^{3}+3\pi\Delta^{4})+6(2\Delta^{6}-2\Delta^{2}-y^{4})X(y)\Big]
\end{equation}
Where, X(y) is an auxiliary function and 
\begin{equation}
y=d \theta/r_{s} \nonumber
\end{equation}
\begin{equation}
\Delta^{2}=1-y^{2}=1-d^{2}\theta^{2}/r_{s}^{2}  \nonumber
\end{equation}

\noindent Due to insufficient kinematics data, the reliability of J-factors for the dSphs and UFDs is thus still under question. However, it is reported by Evans \textit{et al.}, 2016\cite{bib:evan1}, that their formula for J-factor estimation gives more or less accurate results in comparison to the spherical Jeans models driven by Markov Chain Monte Carlo techniques. Evans \textit{et al.}, 2016  \cite{bib:evan1} argued that the analytical formula for J-factors in NFW halo can reproduce the computational results very well. For our purpose, we adopted the J-factors of UMi, Tuc-II and Ret-II from Evans \textit{et al.}, 2016 \cite{bib:evan1}. \\

\subsection{DM annihilation Constraints from Tuc-II}
\subsubsection{\textbf{Searching for $\gamma-ray$ emission due to DM Annihilation from Tuc-II}}

\begin{figure}[h!]
  \subfigure[]
 {\includegraphics[width=0.5\linewidth]{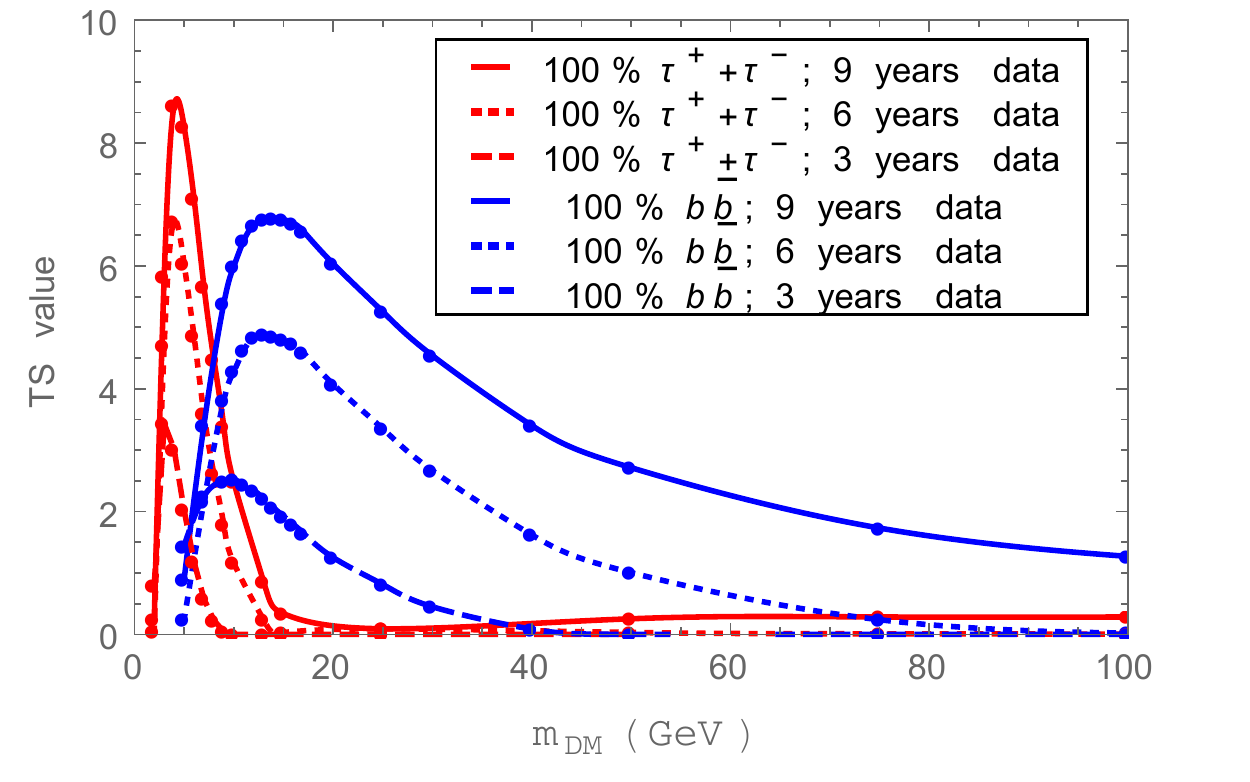}}
\subfigure[]
 {\includegraphics[width=0.5\linewidth]{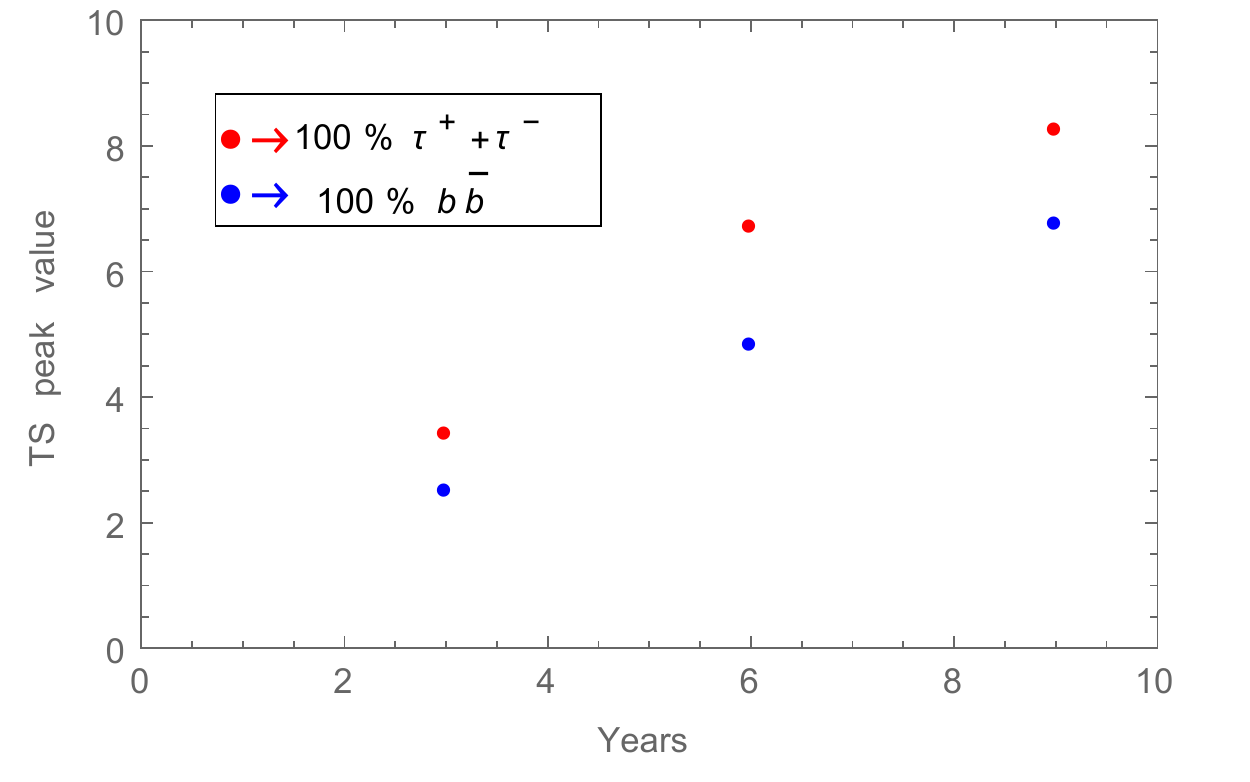}}
\caption{(a) Variation of observed TS values of Tuc-II as a function of DM mass ($m_{DM}$) results from two WIMP pair annihilation channels; $100\%$ $b\bar{b}$ (blue) and $100\%$ $\tau^{+}\tau^{-}$ (red) with three different time periods of \textit{Fermi}-LAT data. (b) The observed TS peak of the excess $\gamma$-ray emission from the direction of Tuc-II for three different time intervals of \textit{Fermi}-LAT data, whereas, the red and blue markers denote the TS peak for $100\%$ $b\bar{b}$ and $100\%$ $\tau^{+}\tau^{-}$ annihilation channels, respectively. The colors and the line-styles of different curves are indicated in the diagram.}
\end{figure}

\noindent In this subsection, we have fitted the possible $\gamma$-ray flux from Tuc-II in terms of the flux arising out of the pair-annihilation of the WIMPs by employing a full-scale MC simulation package DMFit~\cite{bib:jel, bib:gondo}, as implemented in the \texttt{ScienceTools}. The DMFit package is based on the particular set of MC simulations of hadronization and/or decay of the annihilation products as used by the DarkSUSY team~\cite{bib:jel, bib:gondo} by means of the Pythia 6.154~\cite{bib:sjos} event generator. With this $\gamma$-ray spectrum from DM annihilation, we defined Tuc-II as a point source and the significance of the Tuc-II is estimated by the $\Delta TS$ method as we already mentioned in section 2.3. \\

\noindent In Fig.~2(a), we have shown the detection significance of $\gamma$-ray emission, i.e the TS values from the direction of Tuc-II as a function of DM mass ($m_{DM}$) for two pair annihilation channels, $100\%$ $b\bar{b}$ and $100\%$ $\tau^{+}\tau^{-}$. In that same figure, we have also compared the detected TS values from Tuc-II for three, six and nine years of \textit{Fermi}-LAT data and for such purpose we have applied the same analysis method on these three dataset. In Fig.~2(b), we have shown that the TS peak of Tuc-II is increased with the larger dataset and the same nature is followed by both annihilation channels. Even though the observed significance is faint (i.e. less than TS=25) to claim anything strongly, the most encouraging part of this result is that TS peak of Tuc-II is continuously increasing with time and in future this could possibly lead us to a detection of a real signal either from any astrophysical source or from DM annihilation. From Fig.~2(a), we can observe that with nine years of \textit{Fermi}-LAT data, the TS value peaks at $m_{DM}$~=~14 GeV for $100\%$ $b\bar{b}$ annihilation channel, whereas for $100\%$ $\tau^{+}\tau^{-}$ it peaks at $m_{DM}$~=~4 GeV. \\

\noindent There are some studies which have previously analyzed Tuc-II with 6 or 7 years of \textit{Fermi}-LAT data\cite{bib:drlicanew, bib:hoopernew, bib:albert17, bib:calorenew}. In our analysis we have studied it with nine years of \textit{Fermi}-LAT data. The increase in TS values of Tuc-II with 9 years of \textit{Fermi}-LAT data can possibly come from a larger dataset. Thus, this increase in the $\gamma$-ray emission with the largest possible available dataset seems encouraging in indirect detection of DM signal.\\

\begin{table}[h!]
\caption{Summary of test statistics (TS) and $\Delta$ TS for the two source models considered in this paper: power law (PL) for $\Gamma=~2.4$ and the best-fitted dark matter spectrum (DM) corresponds to its highest TS value (in our case $\rm{100\%~\tau^{+}\tau^{-}~channel}$ at DM mass= 4 GeV). Here p-value is derived assuming a $\chi^{2}$ distribution for 1 degree of freedom.}.
\label{Tab-3}
\begin{tabular}{|p{2cm}|p{2cm}|p{2cm}|p{2cm}|p{2cm}|p{2cm}|p{2cm}|p{3cm}|}
\hline 
Our source & TS for PL & $\sigma~(= \sqrt{TS})$ for PL & p-value for PL & TS for DM & $\sigma~(= \sqrt{TS})$ for DM & p-value for DM & $\Delta$ TS (DM-PL)\\
\hline
Tucana-II & 3.59 & 1.89 & 0.05 & 8.61 & 2.93 & 0.003 & 5.02  \\
\hline
\end{tabular}
\end{table}

\begin{figure}
\subfigure[]
 { \includegraphics[width=0.5\linewidth]{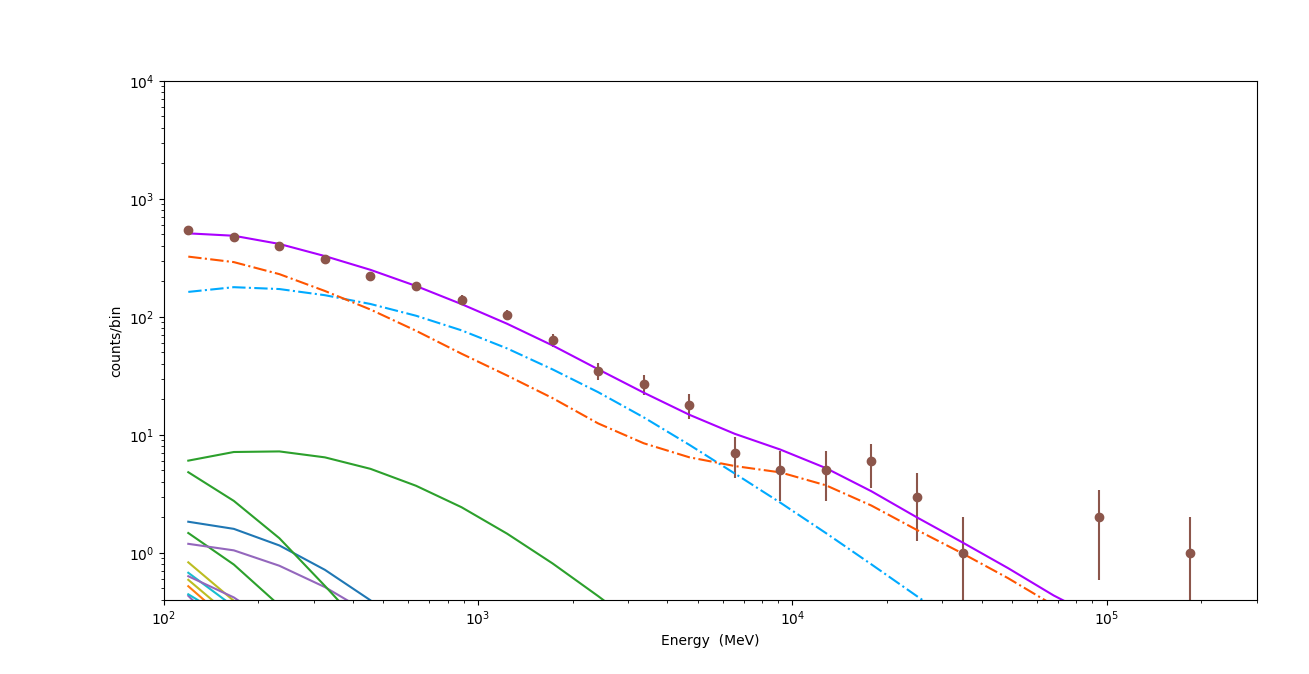}}
\subfigure[]
 { \includegraphics[width=0.49\linewidth]{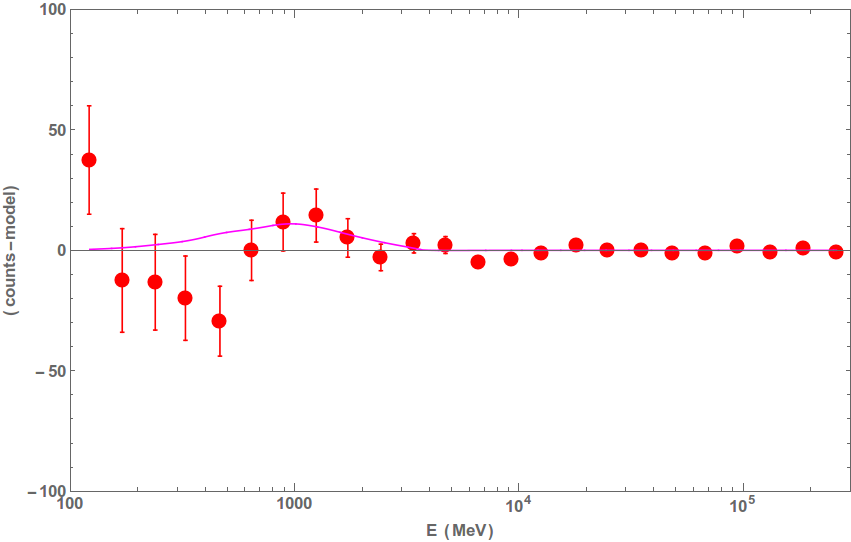}}
\caption{Spectral fit to the counts (Fig.~3(a)) and the corresponding residual plot (Fig.~3(b)) for a $1^{\circ}$ $\times$ $1^{\circ}$ ROI centred on Tuc-II. In Fig. 3(a), the solid purple curve displays the best-fit total spectrum, along with the corresponding LAT-observed data points (in brown); the dot-dashed sky blue and orange curves display the galactic diffuse background and the isotropic background component, respectively. In Fig. 3(b), we show the best-fitted dark matter spectra for 100$\%$ $\tau^{+}\tau^{-}$ annihilation channel at $m_{DM}$ = 4 GeV with a magenta solid line and the residual plot between 100 MeV to 300 GeV energy range is overplotted here as the red points with errorbars.}
\end{figure}

\noindent For both the power-law and the DM annihilation spectra, the best-fitted spectra have been obtained from likelihood ratio test (i.e. $TS = -2\ln\Big(L_{\rm {(max, 0)}}/L_{\rm {(max, 1)}}\Big)$). We have found that the best-fitted TS value is significantly improved with dark matter annihilation hypothesis. Moreover, the p-value (p-value is the probability of getting ``signal-like'' data obtained from the background excess) of local significance is also reduced with DM annihilation spectrum. The p-value is derived by assuming the $\chi^{2}$ distribution for 1 degree of freedom. These details are mentioned in TABLE~5. This table shows that our results may favor the dark matter annihilation hypothesis over its astrophysical connection with the excess obtained from the location of Tuc-II. But it is also important to note that for both power-law (-log(Likelihood) = 11460) and DM annihilation hypothesis (-log(Likelihood) = 11562), we have obtained a comparable -log(Likelihood) value. Therefore, we are not in a position to firmly rule out the astrophysical connection over the DM annihilation hypothesis. Hence we can conclude that our results, at best, show a hint of a DM signal from Tuc-II. For DM model, our obtained $\sigma$ is 2.93 and in the following section we will consider the effect of nearby unresolved sources which has not been taken account into the Fermi-LAT catalog and will study whether such effects will reduce the significance ($\sigma$) for DM annihilation spectrum.\\

\noindent In section 2.3, we have performed the Fermi-LAT analysis on $10^{\circ}~\times~10^{\circ}$ ROI around Tuc-II but with such a large region of the sky, it is quite impossible to identify any interesting features at the location of Tuc-II. So, in Fig.~3(a,b), we have shown the best fit spectra and residual plot for much smaller ROI i.e. for $1^{\circ}~\times~1^{\circ}$ ROI where all the source and background parameters are kept fixed to the best fit values from the $10^{\circ}~\times~10^{\circ}$ ROI fit. In order to investigate any special feature that could originate from the region of Tuc-II, in Fig. 3(a,b), we have shown the data from $1^{\circ}~\times~1^{\circ}$ ROI centered on Tuc-II without the inclusion of Tuc-II in the source model. \\

\noindent Fig.~3(a) shows the spectral fit per energy bin of all the sources within the aforementioned ROI except for Tuc-II, along with the isotropic and the galactic diffuse background components. The corresponding residual plot in the given ROI is shown in Fig.~3(b). In Fig.~3(b), we have also shown the best-fitted DM spectrum of Tuc-II with magenta solid line and the residual plot between 100 MeV to 300 GeV are overplotted with red points. \\

\noindent Here, we have chosen the best-fitted DM spectrum for 100$\%$ $\tau^{+}\tau^{-}$ annihilation channel at DM mass = 4 GeV, which produces the highest TS value of Tuc-II (see Fig.~2(a,b)). In order to quantify the goodness of fit between DM annihilation spectra and data obtained from residual energy spectrum, we have used the T-TEST method which is a frequently used when one is dealing with small number of events. T-TEST is a statistical hypothesis test which indicates whether there is any significant difference between the means of two samples. Under null hypothesis, this test assumes that both the samples are likely to come from same populations (see Appendix A and B). For our analysis, with T-TEST, we have tried to examine whether DM model spectrum can produce an acceptable fit to the data obtained from the residual plot (Fig.~3(b)), where the residuals from all pixels are combined into the energy bins first. In Fig. 3(b), for full energy range (including both positive bump and negative bump at low energy), the dark matter model for $\tau^{+}\tau^{-}$ annihilation channel provides an acceptable fit to the data with a p-value of 0.112 (here p-value is associated with the T-Test goodness of fit). The p-value $>0.05$ indicates that we could not reject the assumption of null hypothesis. Hence, we fail to reject the possibility that the shape of residual energy spectrum is consistent with the DM annihilation spectra(including both positive and negative bumps). Moreover, if we are considering only the positive residual bump between 500 MeV to 5 GeV, the best-fitted DM annihilation spectra produces a good fit to the bump of residual excess with a p-value of 0.782. The positive bump in the residual plot shows then an intriguing hint of possible DM annihilation in Tuc-II. We would also like to note that in Fig 3(b), below 500 MeV, there is a negative bump, roughly as significant as the positive bump in the residual plot. At lower energies, this negative bump may come from the incorrect modeling of the background models. Since our obtained TS values are lower compared to the \textit{Fermi}-LAT threshold detection limit, so we are not in a position to completely eliminate the excess as possible statistical fluctuations or its connection with nearby unmodeled astrophysical sources. In subsection 3.2.2, we discuss this aspect in detail. But our study hints that with more detailed analysis and with a larger data set, Tuc-II could possibly lead us to a detection in DM signal from dSphs. \\

\subsubsection{\textbf{Distribution of the Excess obtained from $\gamma$-ray spectra of DM annihilation}}
\begin{table}[h!]
\caption{A list of BZCAT and CRATES sources that lie within 1$^{\circ}$ of Tuc-II. J225455-592606 is detected in both catalogs, so we listed it with the CRATES coordinates.}
\label{Tab-3}
\begin{tabular}{|p{2cm}|p{6cm}|p{5cm}|}
\hline 
Our source & Nearby sources from BZCAT and CRATES catalog & Distance to the Tuc-II ($^{\circ}$)\\
\hline
Tucana-II & J~225134-580103 & 0.55 \\
\hline
& J~225008-591029 & 0.66 \\
\hline
& J~225455-592606 & 0.95 \\
\hline
\end{tabular}
\end{table}

\begin{table}[h!]
\caption{The summary test statistics (TS) values of Tuc-II, 4FGL 2247.7-5857 and other three nearby sources from CRATES and BZCAT catalog are listed here. For Tuc-II we have chosen the TS peak values for $100\%~\tau^{+}\tau^{-}$ annihilation channel with DM mass=4 GeV and the other three nearby CRATES sources (within $1^{\circ}$ of Tuc-II) are modeled with power-law spectra of $\Gamma=2.2$ \cite{bib:carl}. For 4FGL 2247.7-5857, we have used power-law model and the parameters are taken from Fermi-LAT's 4FGL catalog \cite{bib:4fgl}.}
\begin{tabular}{|p{1cm}|p{2cm}|p{2cm}|p{2cm}|p{2cm}|p{2cm}|p{3cm}|p{3cm}|}
\hline 
Year & Tuc-II from by $\Delta~TS$ method &  TS of J~225134-580103 & TS of J~225008-591029 & TS of J~225455-592606 & TS value of 4FGL 2247.7-5857 & TS value of Tuc-II after including three CRATES sources and 4FGL 2247.7-5857 to source model &  Rescaled TS value of Tuc-II due to all possible background fluctuation\cite{bib:carl, bib:acknew1, bib:albert17}\\
\hline
3 & 3.0868 & 0.05 & 0.027 & 0.49 & 5.61 & 3.04 & $\approx$ 1.7167\\
\hline
6 & 6.8802 & 0.66 & 1.22 & 0.98  & 10.45 & 5.24 & $\approx$ 3.8265\\
\hline
9 & 8.61  & 2.043 & 3.82 & 2.01 & 21.67 & 7.05 & $\approx$ 4.7885\\
\hline
\end{tabular}
\end{table}

\noindent In subsection 2.3 and 3.2.1, we have estimated the TS value by $\Delta~TS$ method but have not checked for any nearby background fluctuation which can also be responsible for the significance obtained from the location of Tuc-II. More importantly, from our analysis, we have obtained a very faint hint of excess from Tuc-II i.e. TS value of 8.61. Hence, in order to claim its connection with DM annihilation, we need to carefully quantify the origin and validity of this excess. \\

\noindent There is a strong possibility that the excess obtained from Tuc-II could come either from the nearby unresolved sources or from the deficiency of background models. Carlson \textit{et al.}, 2015~\cite{bib:carl} have argued that such $\gamma$-ray excess from dSphs can plausibly arise from a number of nearby faint $\gamma$-ray sources including star-forming galaxies \cite{bib:ack1, bib:lind}, radio galaxies \cite{bib:inou}, blazars \cite{bib:abdo1} and millisecond pulsars \cite{bib:hoop1} and a proper multiwavelength study can reduce contamination from these sources. Among all types of different background sources, blazars are the most promising candidates to explain the background fluctuations \cite{bib:carl}. Star-forming and radio galaxies can also provide a non-negligible contribution in $\gamma$-ray sky. But at the high-latitude gamma-ray sky, blazars are the most numerous point sources and they are thought to be the main source of anisotropy in the extragalactic gamma-ray background \cite{bib:ack2, bib:abaza, bib:vent, bib:vent1, bib:cuo, bib:hard}.\\

\noindent Inspired by Carlson \textit{et al.}, 2015~\cite{bib:carl} work, we have tried to perform a more detailed study to investigate the possible reason for obtaining a TS value of $\sim$8.61 from the location of Tuc-II. We have chosen two multiwavelength blazar catalogs i.e. BZCAT \cite{bib:mass} and CRATES \cite{bib:heal}. BZCAT contains nearly 3149 known blazars and 2274 of which are located at high galactic latitude (i.e.$|b|>30^{\circ}$) and CRATES detected more than 11,000 bright flat-spectrum radio sources. Within $1^{\circ}$ of Tuc-II, we have found three radio sources from the CRATES catalog and one blazar from the BZCAT catalog. The source from BZCAT catalog has also been detected by CRATES. For our investigation, we have only considered CRATES sources (J225134-580103, J225008-591029, and J225455-592606), which are located within a 1$^{\circ}$ from Tuc-II because any blazars, radio or star-forming region beyond 1$^{\circ}$ possibly would not produce any effective changes to the local significance of dSphs \cite{bib:carl}. The list of CRATES sources within 1$^{\circ}$ of Tuc-II is mentioned in TABLE~6.\\

\noindent We have modeled these three sources with the power-law spectrum of $\Gamma$=2.2 \cite{bib:carl} and have determined the TS values of these sources for three different time periods of Fermi-LAT data. The result of this analysis is that by including these three sources, the significance of Tuc-II is only decreased by $\sim$ 10$\%$ (mentioned in Table~7). We wish to note that Carlson \textit{et al.}, 2015~\cite{bib:carl}, has also obtained the same. They have also concluded that blazars are only responsible for only about 10$\%$ of actual TS value of the source and the major portion of the source excess is unlikely to be related to the nearby BZCAT and CRATES sources.\\ 

\noindent In order to check the distribution of excess obtained from Tuc-II, we have created the $2^{\circ}$ x $2^{\circ}$ residual TS map (100 MeV - 300 GeV) around Tuc-II with \textit{`gttsmap'}. In this process, the model parameters of all the sources within ROI of $10^{\circ}~\times~10^{\circ}$ were kept fixed to their best-fitted values obtained from the binned likelihood analysis on 9 years of Fermi-LAT data and the normalization parameters of both the galactic and isotropic components were left free. We have run the same process for three cases: In Fig.~4(left); we have not included Tuc-II and three nearby BZCAT and CRATES sources, that lie within a $1^{\circ}$ of Tuc-II, to the source model, in Fig.~4(middle); we have included the three nearby radio BZCAT and CRATES sources to source model but not the Tuc-II, in Fig.~4(right); we have included the Tuc-II and also three BZCAT and CRATES sources to the source model. Here, for Tuc-II the best fitted parameters obtained from dark matter annihilation spectra (i.e. $100\%~\tau^{+}\tau^{-}$ channel at DM mass=4 GeV) has been used to generate the third residual TS map. \\

\begin{figure}
 { \includegraphics[width=1.0\linewidth]{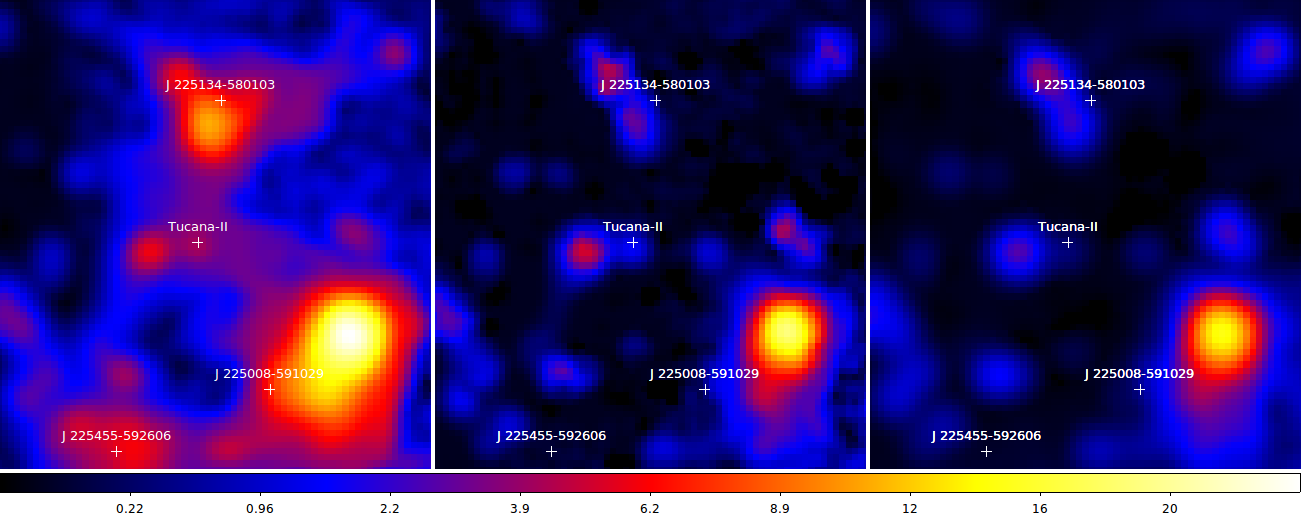}}
\caption{ Residual TS maps (100 MeV - 300 GeV) of $2^{\circ}~\times~2^{\circ}$ regions centered at Tuc-II extracted from the $10^{\circ}$ $\times$ $10^{\circ}$ ROI data around Tuc-II. The image scale of TS map is 0.025 $pixel^{-1}$. In left Fig., Tuc-II and the three CRATES sources are not included in the source model; in middle Fig., the three CRATES sources are included in the source model but not the Tuc-II; in right Fig., Tuc-II and the three CRATES sources have been included in the source model. The name of all three CRATES sources within a $1^{\circ}$ from Tuc-II and the Tuc-II are mentioned with a white cross.} 

\end{figure}

\noindent From the Fig.~4(left, middle), we find that there is a hint of the localized-excess of TS value around $\approx$ 6.5 which is very close to the Tuc-II location. This excess is not exactly localized to the position Tuc-II, but it is just $0.18^{\circ}$ away from the Tuc-II. In Fig.~4(right), after including Tuc-II as well as three CRATES sources to the source models, the significance of the excess-region near Tuc-II is significantly reduced. From Fig.~4, we can then state that there is a possibility that the nearby localized excess-region might associate with Tuc-II.\\

\noindent From Fig. 4, even after including the blazars in the source model, we can observe a bright source of $\approx$5$\sigma$ just at the bottom of the right corner of the TS map. We have checked the very recently published 4FGL catalog of Fermi-LAT (\cite{bib:4fgl}) and have found that a new source, namely 4FGL 2247.7-5857, is exactly overlapping with that excess region. Hence, we have again generated the residual TS map of $2^{\circ}~\times~2^{\circ}$ for four cases (Fig. 5); the first and the second residual TS maps in Fig. 5 are the same as the first two residual TS maps from Fig. 4, in the third TS map of Fig. 5; along with three CRATES sources we have now included the 4FGL 2247.7-5857 to our source model but not the Tuc-II, in the fourth (Fig.~5(right)); we have included the Tuc-II, three BZCAT and CRATES sources and also the 4FGL 2247.7-5857 to the source model. \\

\begin{figure}
 { \includegraphics[width=1.0\linewidth]{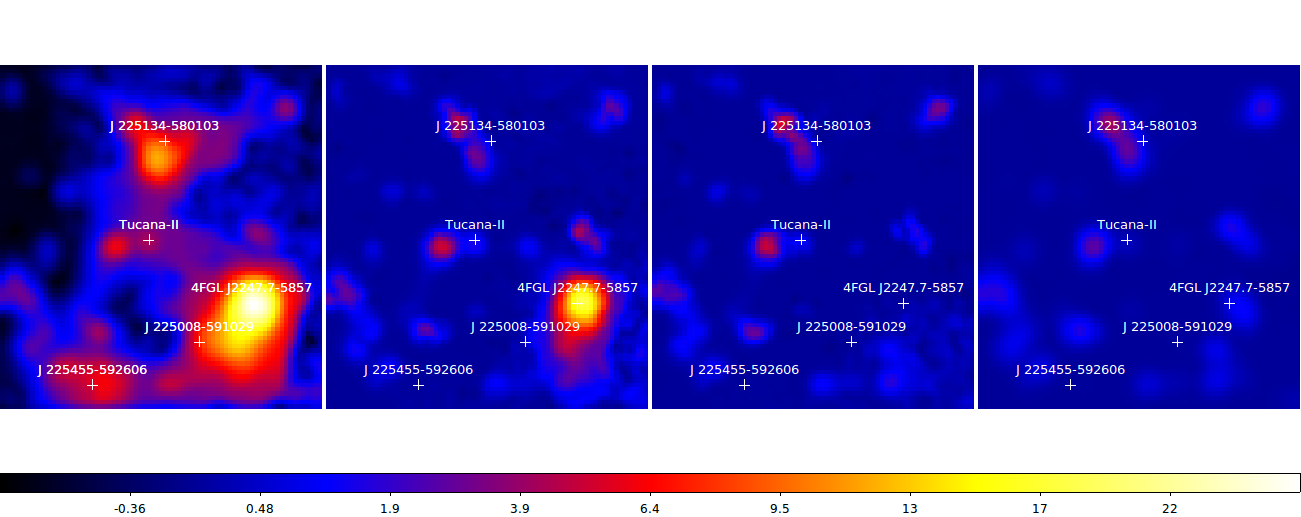}}
\caption{ Residual TS maps (100 MeV - 300 GeV) of $2^{\circ}~\times~2^{\circ}$ regions centered at Tuc-II extracted from the $10^{\circ}$ $\times$ $10^{\circ}$ ROI data around Tuc-II. The image scale of TS map is 0.025 $pixel^{-1}$. In First Fig., Tuc-II, the three CRATES sources and 4FGL 2247.7-5857 are not included in the source model; in Second Fig., the three CRATES sources are included in the source model but not the Tuc-II and 4FGL 2247.7-5857; in Third Fig., the three CRATES sources and 4FGL 2247.7-5857 have been included in the source model but not the Tuc-II; in Fourth Fig., Tuc-II, three CRATES sources and 4FGL 2247.7-5857 have been included in the source model. The name of three CRATES source, 4FGL 2247.7-5857 and Tuc-II are mentioned with a white cross.} 

\end{figure}

\noindent One can see from the extreme right panel of Fig. 5 that after including 4FGL 2247.7-5857 to the source model, the bright excess at the bottom of the right corner is significantly reduced. It shows that the bright excess from the right-bottom of TS map has an astrophysical connection and it mainly comes from the source 4FGL 2247.7-5857. We again observe that after including Tuc-II to source model, the significance of the nearby localized excess (i.e. $0.18^{\circ}$ away) of Tuc-II is considerably decreased (see right most panel of Fig.~5).\\

\noindent We would like to point out that, even after including three CRATES sources and 4FGL 2247.7-5857, from both the residual TS maps (Figs.~4 and 5), we can still observe a number of delocalized regions of excess. This may be due to the deficiency of the current background models of Fermi-LAT. There is a fair chance that this delocalized excess might come from unresolved astrophysical sources but we also should not ignore the possibility of presence of dark matter subhalos. Some studies argue that even if all the astrophysical sources are accurately modeled, the dark matter subhalos will still be responsible for an irreducible background ($\approx5\%-10\%$) for gamma-ray sky \cite{bib:carl, bib:acknew2, bib:lee1, bib:siegal}. We can expect that in future with detailed multiwavelength study, it would be possible to reduce the contamination from unresolved sources in the blank sky.\\

\noindent In our study, we have calculated the TS value only with respect to Fermi-LAT provided background model and not from the blank sky location. So, there is a fair possibility that we have overestimated the TS value even after including all the known sources to the source model. Several Fermi collaboration papers \cite{bib:acknew1, bib:albert17} already report that in a large number of blank sky positions, excess of TS $>$ 8.7 is very common. This decreases the significance from 2.95$\sigma$ to 2.2$\sigma$ \cite{bib:acknew1, bib:albert17}. Using this prescription \cite{bib:acknew1, bib:albert17}, we have re-calibrated our obtained TS values and it reduces the TS value of Tuc-II from 8.61 (p value= 0.003) to 4.79 (p value= 0.029). All these results are shown in TABLES~7. In column 2, we have given our obtained TS value from $\Delta TS$ method; in columns 3, 4, 5 and 6 the TS value of all three CRATES sources and 4FGL 2247.7-5857 are mentioned; in column 7, we have shown the modified TS value of Tuc-II after including three CRATES sources and 4FGL 2247.7-5857 to the model; and in column 8, the re-calibrated TS value of Tuc-II for all possible background fluctuations has been shown.\\

\subsubsection{\textbf{Possible DM annihilation Constraint on theoretical DM Models with 9 years of Tuc-II \textit{Fermi}-LAT data}}
\noindent Since with dark matter annihilation spectra, our obtained TS peak value for $\tau^{+}\tau^{-}$ annihilation channel is weaker than Fermi-LAT' threshold detection limit (i.e. TS~$<$~25), in this section we estimated $\gamma$-ray flux upper limit in $95~\%$ C.L. from Tuc-II by semi-Bayesian method~\cite{bib:helene}, as described in subsection 2.3 of this paper. By employing the $\gamma$-ray spectrum from DM annihilation, we could also calculate the upper limits to the thermally averaged pair-annihilation $<\sigma v>$ of the WIMPs with the variation of the plausible WIMP masses ($\rm{m_{DM}}$), for various important pair-annihilation final states (f), i.e. for each of the possible important channels in which WIMP annihilations might take place to produce $\gamma$-rays~\cite{bib:jung}. We have considered five supersymmetry-motivated pair annihilation final states \cite{bib:jung}, namely $100\%$ $b\bar{b}$, $80\%$ $b\bar{b}$+$20\%$ $\tau^{+}\tau^{-}$, the $100\%$ $\tau^{+}\tau^{-}$, $100\%$ $\rm{\mu^{+} \mu^{-}}$ and $100\%$ $W^{+}W^{-}$, respectively. The variation of such 95 $\%$ $\gamma$-ray flux upper limits of Tuc-II obtained from semi-Bayesian method and the relative upper limits to their annihilation $<\sigma v>$ with increasing WIMP masses are displayed in Fig.~6(a,b), separately for each of the annihilation channels mentioned above.\\

\begin{figure}[h!]
\subfigure[]
 {\includegraphics[width=0.45\textwidth,clip,angle=0]{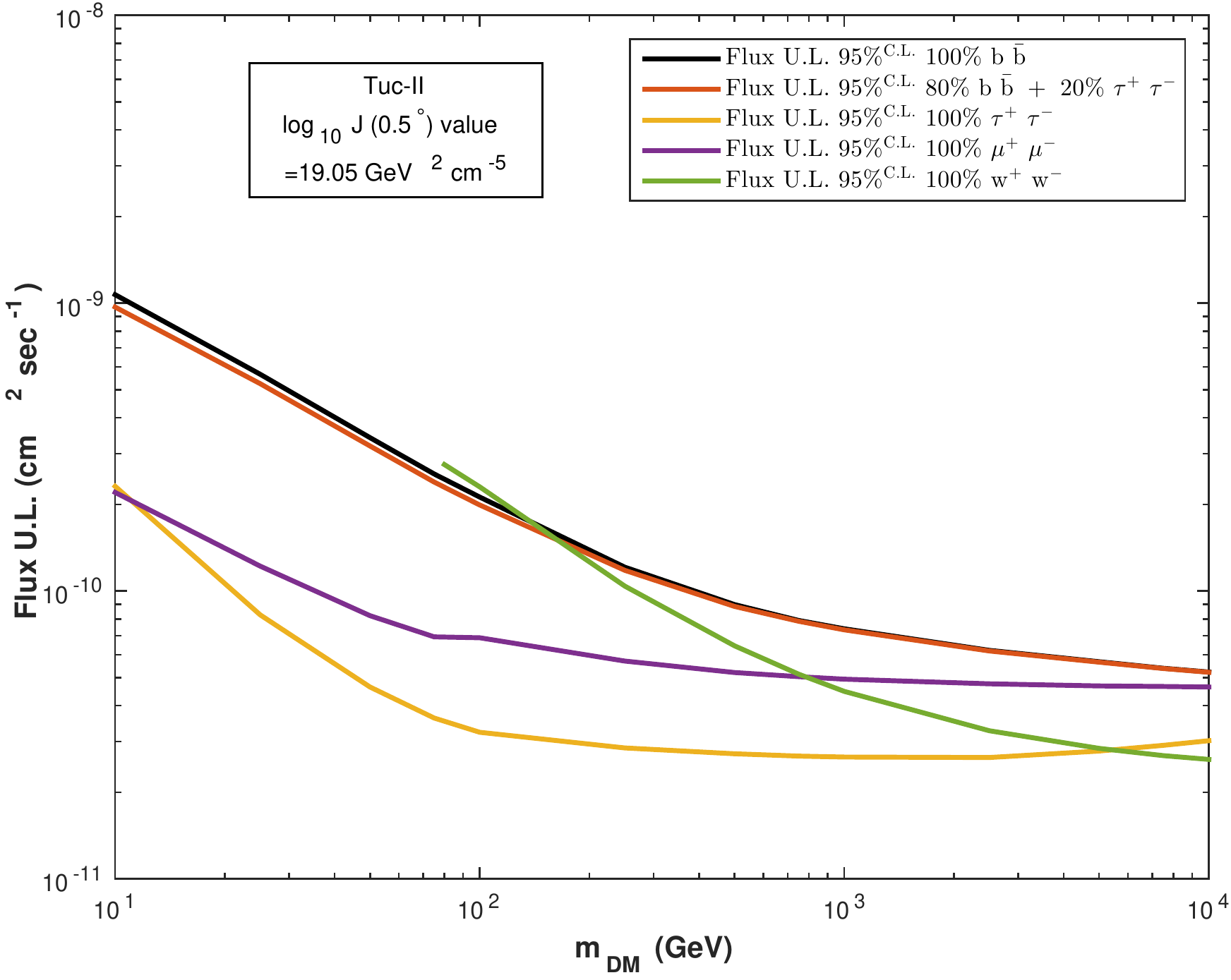}}
\subfigure[]
 { \includegraphics[width=0.45\linewidth]{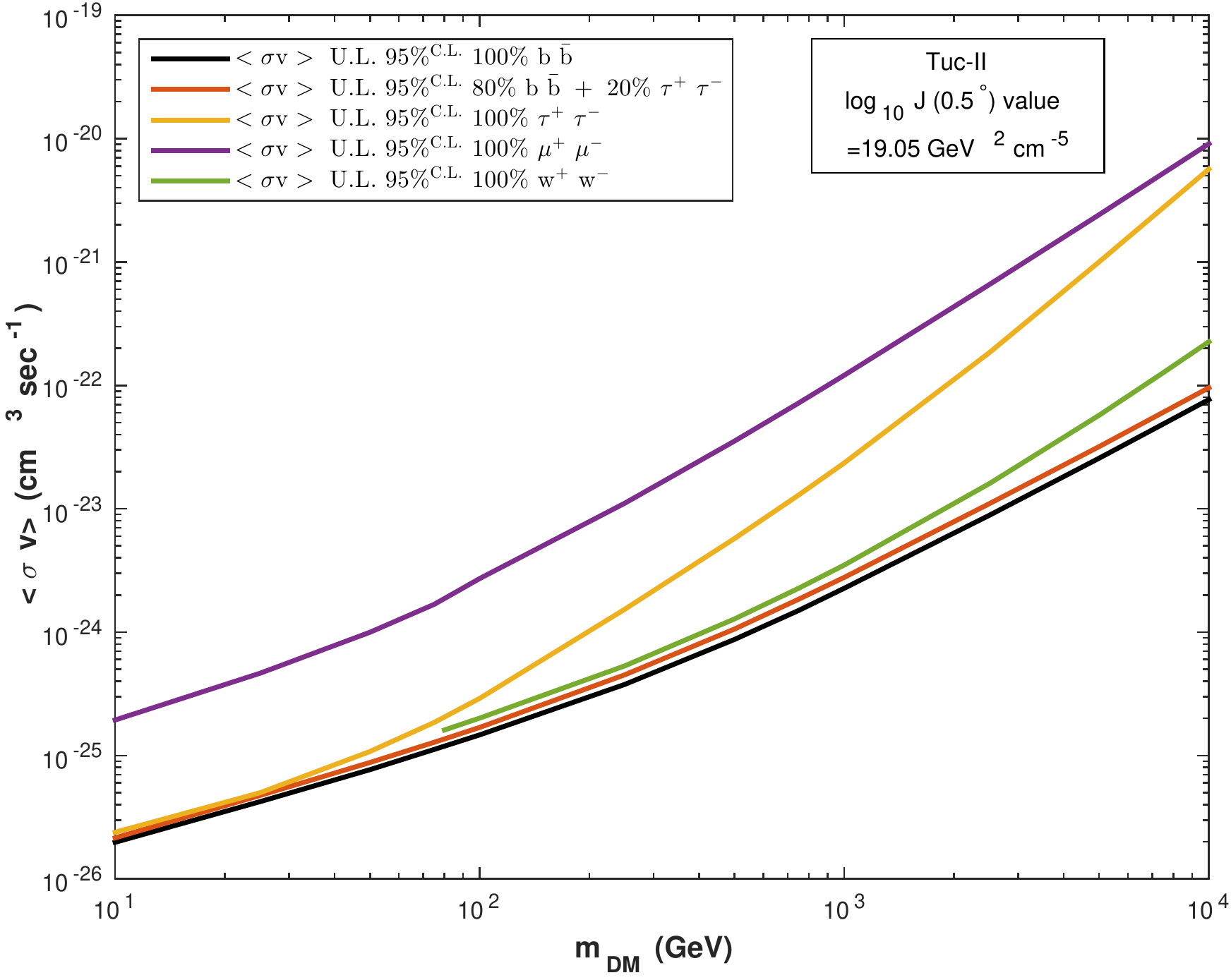}}
\caption{Variations of (a) the $95\%$ C.L. $\gamma$-ray (integral) flux upper limits of Tuc-II and (b) the corresponding pair-annihilation $<\sigma~v>$ of the WIMPS with their increasing masses ($\rm{m_{DM}}$), as estimated for various annihilation final states ``f" (indicated in the diagram) by using the DM annihilation function with semi-Bayesian likelihood method. Each of these results is estimated for the median $\rm{J(0.5^{\circ})}$-factor value of Tuc-II\cite{bib:evan1}.}
\end{figure}

\begin{figure}[h!]
\subfigure[]
 { \includegraphics[width=0.48\linewidth]{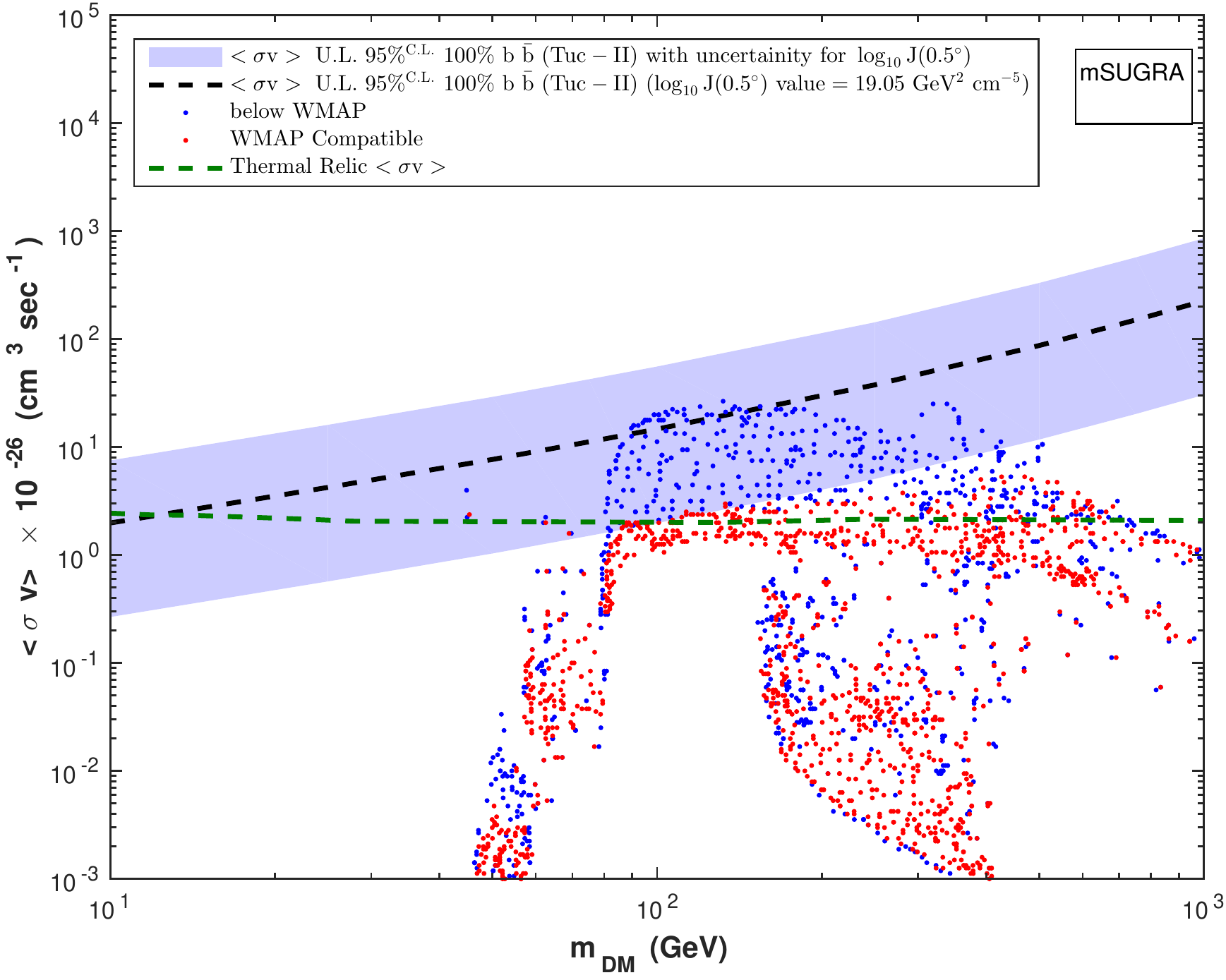}}
\subfigure[]
 { \includegraphics[width=0.48\linewidth]{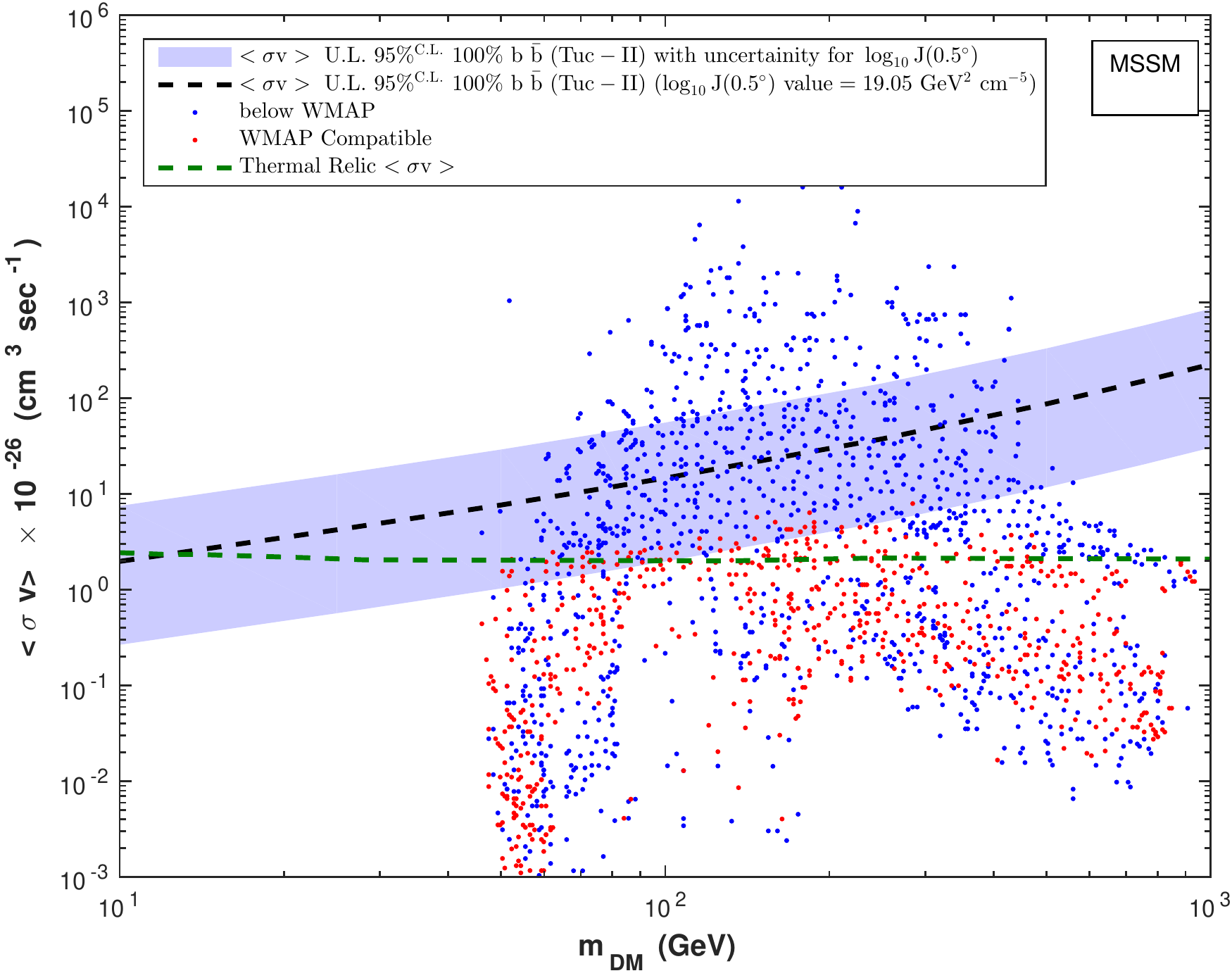}}
\caption{Variation of $95\%$ C.L. upper limit of the WIMP pair-annihilation $<\sigma v>$ with increasing $\rm{m_{DM}}$ for $b\bar{b}$ annihilation final states of Tuc-II displayed in ($\rm{m_{DM},<\sigma v>}$) plane for J-factor with its uncertainties. The shaded band represents the uncertainty in the DM density profiles in the Tuc-II. In the figures, the $<\sigma~v>$ are compared with points derived from (a) the mSUGRA and of (b) the MSSM models~\cite{bib:abdo}. In those later models, the red points correspond to thermal relic density compatible with the WIMP data. The blue points represent higher $<\sigma~v>$, and correspondingly lower thermal relic densities, obtained by assuming certain additional nonthermal production mechanisms to contribute to WIMP production, while the WIMPs still comprise all of the DM. In both the figures, we have also overplotted the relic abundance (or thermal) cross section ($\rm{2.2\times10^{-26}~cm^{3}~s^{-1}}$) estimated by Steigman \textit{et al.}, 2012~\cite{bib:steigman} and it is displayed as green dashed line.}
\end{figure} 
\begin{figure}[h!]

 {\includegraphics[width=0.5\linewidth]{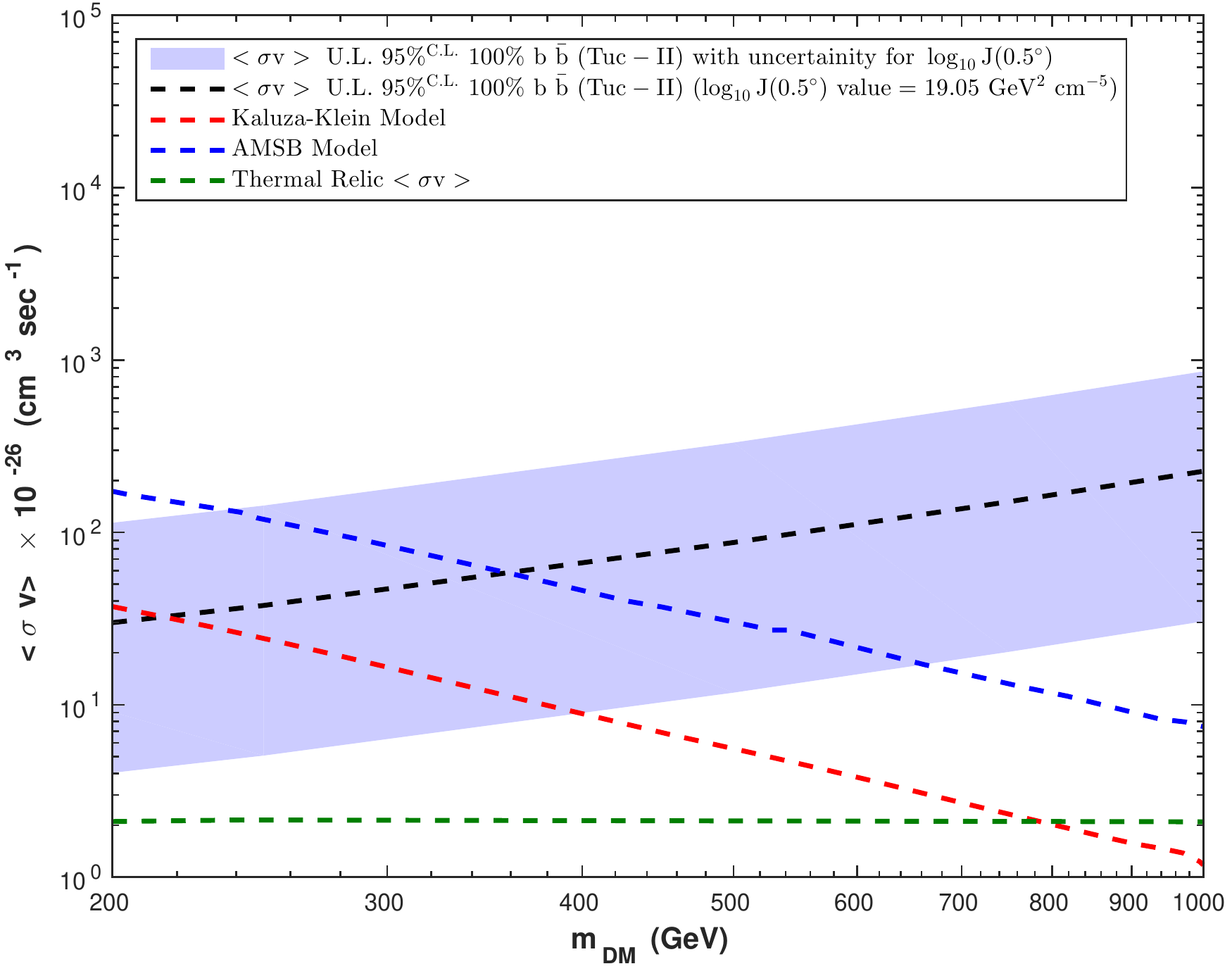}}
\caption{A comparison between the $\gamma$-ray spectrum from DM annihilation for $95\%$ C.L. upper limit of $<\sigma~v>$ vs. $\rm{m_{DM}}$ curve from Tuc-II and the theoretical $<\sigma~v>$ vs. $\rm{m_{DM}}$ curves obtained from the AMSB and the Kaluza-Klein UED models. The shaded band represents the uncertainty in the DM density profiles in the Tuc-II for $b\bar{b}$ annihilation final states. We have also overplotted the relic abundance (or thermal) cross section ($\rm{2.2\times10^{-26}~cm^{3}~s^{-1}}$) estimated by Steigman \textit{et al.}, 2012 \cite{bib:steigman} and it is displayed as green dashed line.}
\label{fig:fig}
\end{figure}

\noindent In Figs.7~(a,b) and 8, we have displayed the DM annihilation function-determined $95\%$ C.L. upper limit of the thermally averaged WIMP pair-annihilation $<\sigma v>$, as a function of the WIMP mass ($\rm{m_{DM}}$) for its median J-factor and J-factor uncertainties \cite{bib:evan1}. Only the $<\sigma~v>$ upper limit in $100\%$ b$\rm{\bar{b}}$ annihilation channel has been considered in these figures as they are found to put the most stringent limits on the parameter space of the models. In Figs.~7(a,b) and 8, the horizontal dashed green line denotes to the relic abundance (or thermal) cross section estimated by Steigman et. al.~\cite{bib:steigman}. These results are then compared with the $<\sigma~v>$ values obtained for various WIMP masses ($\rm{m_{DM}}$) from four theoretical DM models, namely the minimal Supergravity (mSUGRA; in Fig.~7(a))~\cite{bib:cham} model, the Minimal Supersymmetric Standard Model (MSSM; in Fig.~7(b)) \cite{bib:chu}~model, the Anomaly Mediated Supersymmetry Breaking model (AMSB) \cite{bib:giu} (in Fig.~8) model and the lightest Kaluza-Klein particle of Universal Extra Dimensions (UED) model \cite{bib:che, bib:ser, bib:hoop} (in Fig.~8), respectively.\\

\noindent In mSUGRA model, the supersymmetry breaking parameters are defined at the order of grand unification scale $\sim 2 \times 10^{16}$~GeV i.e. at high energy scale, whereas, for MSSM model, the supersymmetry breaking parameters are defined at the low energy scale i.e. at electroplate scale. For AMSB model, the supersymmetry breaking parameters are considered to produce winos; these winos or wino-like neutralino basically are the supersymmetric fermionic partner of the gauge bosons from Standard Model. At about $2$ TeV wino mass, their thermal relic density matches with the universal DM density, on the other hand, several non-thermal DM production mechanisms can explain the winos with comparatively less massive DM particles\cite{bib:abdo}. In Kaluza-Klein model, its first order excitation term of U(1) hypercharge gauge boson is connected to the DM candidate and at about $700$~GeV DM mass, this model can define the thermal relic abundance from their DM density. In Figs.~7(a,b) the blue points in both of the models represent the low thermal relic density with additional nonthermal production mechanisms for the WIMPs to describe the universal matter density, on the other hand, the red points are consistent with the cosmological thermal relic density \cite{bib:abdo}.\\

\noindent From the Figs.7~(a,b) and 8, it is immediately evident that lowest limit of the shaded band of Tuc-II would impose a very strong constraint on the parameter spaces of the popular theoretical WIMP models. In Figs.~7(a,b), it is interesting to note that, for the median J($0.5^{\circ}$)-factor of Tuc-II (i.e. for $\rm{\log_{10} J(0.5^{\circ})}$=19.05~$\rm{GeV^{2}~cm^{-5}}$), the upper limits of $<\sigma~v>$ considerably constrain the blue points in both MSSM and mSUGRA model, while the J-factor uncertainty band of Tuc-II have already begun to constrain the red points in both the models. Fig.~8 shows that the upper limit of $<\sigma v>$ from Tuc-II for the median J($0.5^{\circ}$)-factor (i.e. for $\rm{\log_{10} J(0.5^{\circ})}$=19.05~$\rm{GeV^{2}~cm^{-5}}$), as obtained from Eq.~(6) above disfavors the AMSB and the Kaluza-Klein UED models for masses $\approx<400$~GeV and $\approx<220$~GeV respectively.\\

\noindent The large uncertainties in J-factor of Tuc-II comes from its insufficient kinematics data. With more precise observation of the internal structure of Tuc-II, in future we should definitely reduce this uncertainty band to a possible single upper limit curve of $<\sigma~v>$ and that might improve the constraint limit on beyond Standard Model. This result would then possibly signify the hint of new physics in the field of indirect dark matter detection.

\subsubsection{\textbf{Comparison of the constraints on the DM annihilation cross section ($b\bar{b}$ channel) obtained from Tuc-II, Ret-II and UMi}}

\begin{figure}[h!]
\begin{center}
\subfigure[]
 {\includegraphics[width=0.5\textwidth,clip,angle=0]{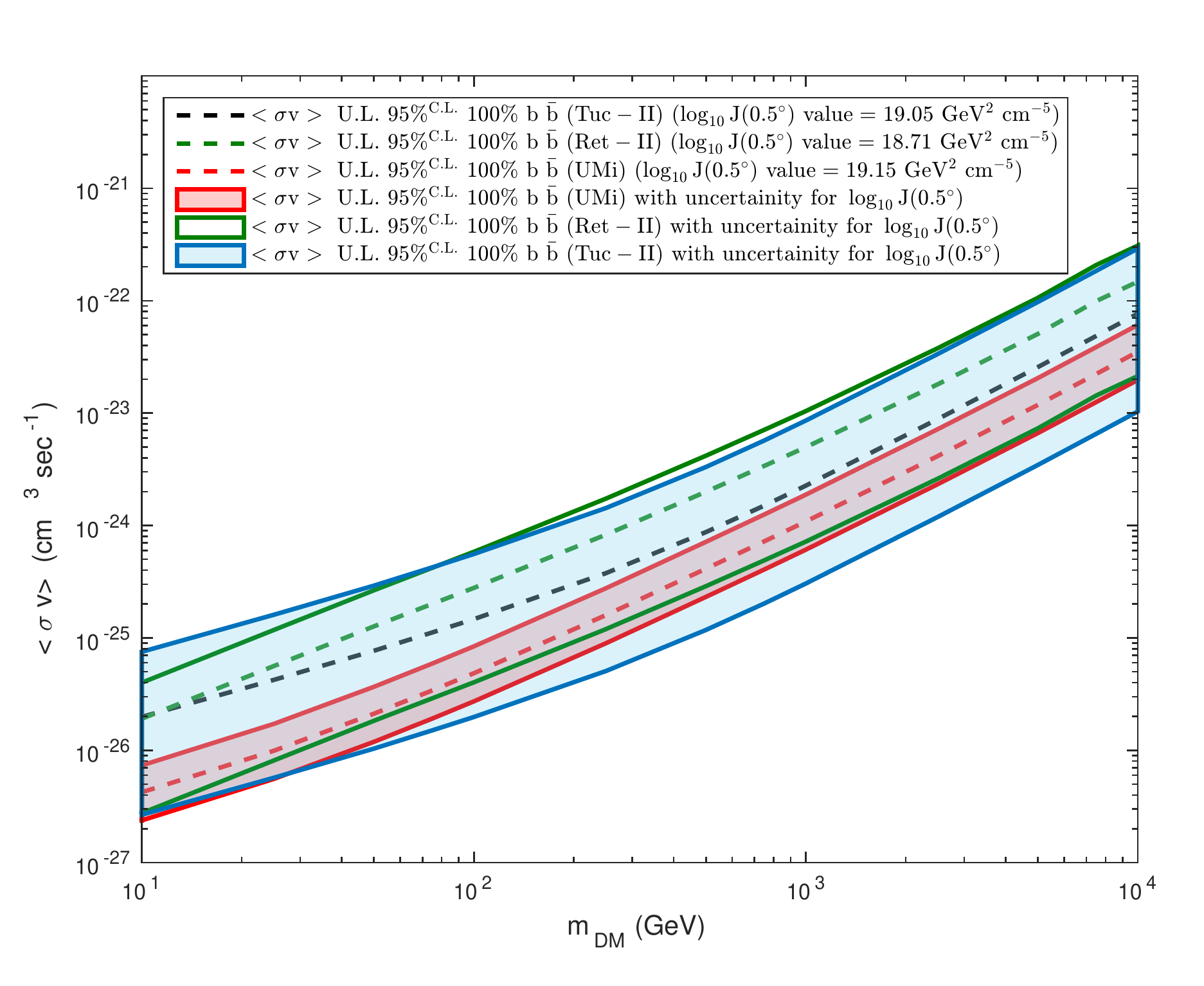}}
\caption{Variations of (a) $95\%$ C.L. upper limit of the WIMP pair-annihilation $<\sigma v>$ with increasing $\rm{m_{DM}}$ for $b\bar{b}$ annihilation final states of Tuc-II, Ret-II and UMi displayed in ($\rm{m_{DM},<\sigma v>}$). The shaded bands represent the uncertainty in the DM density profiles of UFDs and the dashed line denotes the upper limit of $<\sigma v>$ in 95 $\%$ C.L. corresponds to median J-factor \cite{bib:evan1}. The colors and the line-styles of different curves are indicated in the diagram.}
\end{center}
\end{figure}

\noindent In this section, we have compared the upper limit of $<\sigma~v>$ obtained from Tuc-II with two other UFDs, namely UMi and Ret-II, respectively. In Fig.~9 we have shown the $95\%$ C.L. upper limit of the WIMP annihilation $<\sigma~v>$ for only $b\bar{b}$ channel obtained by the analysis of nine years of LAT-data of Tuc-II, UMi and Ret-II. For estimating the 95$\%$ C.L. $<\sigma~v>$ upper limit of these two dwarfs, we have performed the same analysis method that we have applied for Tuc-II (mentioned in TABLE~2).\\

\noindent In Figs.~9 the dashed lines correspond to the median value of J-factor, while the shaded regions depict the range of uncertainty in J-factors of each of the UFDs. In the case of UFDs, very few numbers of stars have been detected so far which is the main obstacle in understanding the DM distribution in UFDs. The large uncertainty bands in Tuc-II basically represent our inadequate knowledge of its internal structure.   \\

\noindent It may be of some interest to note that, in case of Ret-II, several recent studies have detected a slight excess in $\gamma$-emission \cite{bib:drlicanew, bib:hoopernew, bib:geringernew2, bib:albert17, bib:linew, bib:zhaonew}. Though the significance of excess emission from Ret-II is quite lower than \textit{Fermi}-LAT's threshold value for detection, this excess emission is suspected to be the evidence of WIMP annihilation in Ret-II  \cite{bib:geringernew1, bib:linew, bib:albert17, bib:zhaonew}.\\

\noindent Compare to Ret-II and UMi, Tuc-II indicates larger uncertainties in DM density profile and from Fig.~9, we could distinctly observe an overlapping region among the Ret-II, Tuc-II and UMi in parameter space of ($<\sigma~v>$, $m_{DM}$). So, in view of the indirect DM search, it is not possible to favor Tuc-II over other two dSphs but from Fig.~9, it is also evident that above $m_{DM}~\sim~100~GeV$, Tuc-II provides a better limits on ($<\sigma~v>$, $m_{DM}$) space than Ret-II for their median J($0.5^{\circ}$) values.\\

\section{Conclusions and Discussion}
\noindent In this paper, we have analyzed nearly nine years of \textit{Fermi}-LAT $\gamma$-ray data from one of the recently discovered UFDs, namely Tuc-II, with the aim of detecting the signatures of self-annihilation of the WIMPs that are usually believed to be the constituent particles of DM within the UFDs/dSphs. We have found a very faint $\gamma$-ray emission from Tuc-II with both Power law and $\gamma$-ray spectrum from DM annihilation. It is interesting to note that, with $\gamma$-ray spectrum from DM annihilation, we have shown the variation of detected TS values from Tuc-II for various DM masses and have found that with nine years of \textit{Fermi}-LAT data TS value peaks at ~$m_{DM}\sim$14 GeV for $100\%$ $b\bar{b}$ annihilation channel, whereas for $100\%$ $\tau^{+}\tau^{-}$ it peaks at $m_{DM}\sim$4 GeV. In our Galactic Center, the $m_{DM}$ between 25 GeV - 70 GeV for $b\bar{b}$ annihilation channel and $m_{DM}$ between 8 GeV - 15 GeV for $\tau^{+}\tau^{-}$ annihilation channel play an important role to interpret the $\gamma$-ray emission resulting from WIMP annihilation \cite{bib:gordon, bib:hoop2, bib:daylan, bib:zhou, bib:cal}. In our analysis for Tuc-II, the TS peaks obtained for the $m_{DM}$ for two annihilation channels are slightly lower than the DM mass range needed to explain the DM interpretation in Galactic Center. \\

\noindent We have also shown that this excess is increased for larger periods of data and that increase in source significance for TS peak value is roughly proportional to $\sim\sqrt{t}$ \cite{bib:Charles}; where t is the time periods of data. The most interesting part of the result is that the cumulative increase in TS peak values of Tuc-II with larger periods of data which can possibly hint at presence of any real signal either from an astrophysical source or from DM pair annihilation. In indirect DM detection, the hints of increasing $\gamma$-ray excess from some particular dSphs (Tuc-II for our case) may lead us to a new direction of DM physics.  \\ 

\noindent By assuming the $\gamma$-ray spectra for DM annihilation to 100$\%$ $\tau^{+}\tau^{-}$ channel, from Tuc-II location we have obtained a p-value $\approx$ 0.003 with respect to the Fermi-LAT provided background models and it may also come from the rare statistical fluctuation in background. One of the most tantalizing explanations of such excess is the presence of any nearby unresolved bright sources. Among all types of unresolved background sources, blazars are assumed to be the most likely candidates to emit $\gamma$-ray emission just below the Fermi-LAT's threshold detection. Searching from the BZCAT and CRATES catalog, we have found that there are three radio sources within $1^{\circ}$ of Tuc-II and amongst all the most nearby source (J225455-592606) lie at $0.55^{\circ}$ away from Tuc-II. We have also checked the recently published 4FGL catalog of Fermi-LAT (\cite{bib:4fgl}) and have found that a new source, namely, 4FGL 2247.7-5857 which is 0.66 degree away from Tuc-II location. Thus it makes very unlikely that the excess from Tuc-II location would be highly contaminated by these sources. \\

\noindent In Figs.~4 and 5, we have shown the TS maps for energy $>$ 100 MeV. From these TS maps, we find an excess of TS value $\approx$ 6.5 which is just $0.18^{\degree}$ from the Tuc-II. We have also noticed that whenever we have included the Tuc-II to the source model, the emission from that excess region is significantly reduced. Hence, there is a very high possibility that this emission is associated with Tuc-II. All our maps have been generated at energies $>$ 100 MeV. Since at lower energies the point spread function (PSF) of Fermi-LAT is relatively large, but at higher energies ( for example at 500 MeV), the 68$\%$ of the photons will lie within 1 degrees of the source location \footnote{\url{http://www.slac.stanford.edu/exp/glast/groups/canda/lat_Performance.htm}}, we have generated a TS map for energy $>$ 500 MeV. Interestingly, the TS map (see Fig 10) shows that after including Tuc-II to the source model, the nearby excess region has almost disappeared. This probably means that in Figs.~4 and 5 the nearby residual excess emission even after including Tuc-II to source model is related to the poor background modelings. Hence, from our result, we might conclude that the very nearby excess region is connected with Tuc-II and at best, it may show a hint of a DM signal from Tuc-II.\\

\noindent Several Fermi collaboration papers observe that in a large region of blank sky, the excess of TS $>$ 8.7 is very common. If we consider the nearby blazars, they will only account for 10$\%$ of these excesses. The dark matter subhalos may also be responsible for a $\approx$5$\%$-10$\%$ irreducible background. Therefore, we have re-calibrated our obtained significance and it reduces the TS value of Tuc-II from 8.61 (p value=0.003) to 4.79 (p value=0.029). At present, with nine years of data, the obtained emission from Tuc-II is much weaker than Fermi-LAT's threshold detection. But from our work, we have also found that the significance of Tuc-II is increased with increase in time periods of data and from TS map we have also observed a localized excess just beside the Tuc-II. So, in future, with even more time periods of data and with better background modeling, we can expect to positively explain the $\gamma$-ray excess emission from the direction of Tuc-II.\\

\begin{figure}
 { \includegraphics[width=.8\linewidth]{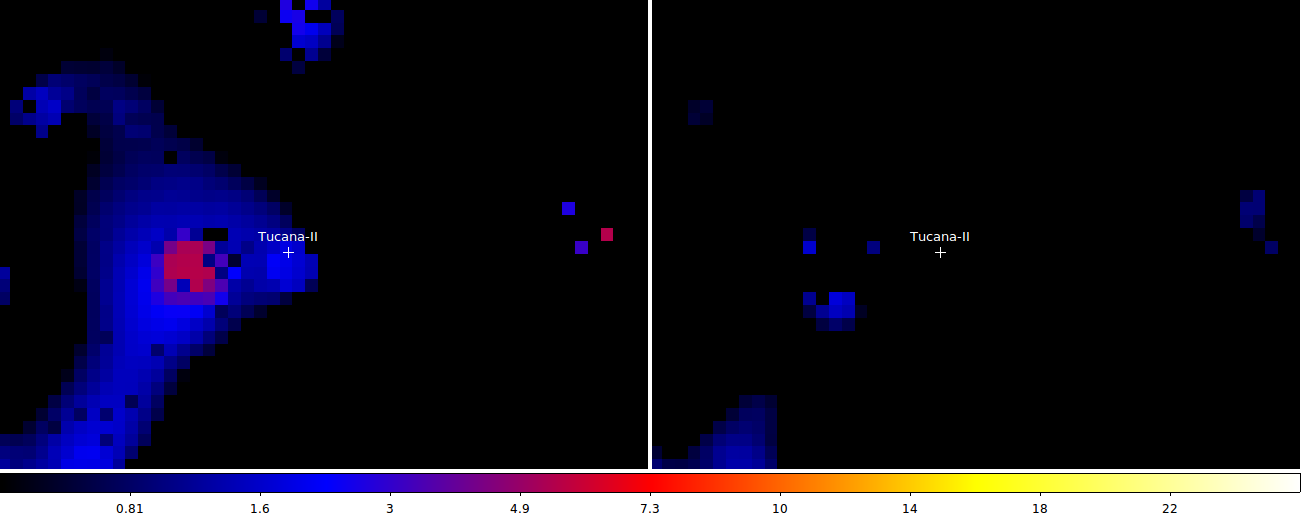}}
\caption{ Residual TS maps (500 MeV - 300 GeV) of $1^{\circ}~\times~1^{\circ}$ regions centered at Tuc-II extracted from the $10^{\circ}$ $\times$ $10^{\circ}$ ROI data around Tuc-II. The image scale of TS map is 0.025 $pixel^{-1}$. In left Fig., the three CRATES sources and 4FGL 2247.7-5857 are included in the source model but not the Tuc-II; in right Fig., Tuc-II, the three CRATES sources and 4FGL 2247.7-5857 have been included in the source model.} 

\end{figure}

\noindent Since the excess obtained from Tuc-II location is below the Fermi-LAT's detection level, we have then calculated the possible upper-limit of the pair-annihilation $<\sigma~v>$ of the WIMPs in Tuc-II, as the function of WIMP mass for five annihilation channels. This method assumes that the entire $\gamma$-ray emission arises for particle interaction in that channel only. We have adopted the J-factor from Evans \textit{et al.}, 2016 \cite{bib:evan1} (Eq.~(6)). \\

\noindent In this paper, we have used larger periods of data compared to the other studies which have analyzed Tuc-II previously \cite{bib:drlicanew, bib:hoopernew, bib:albert17, bib:calorenew} and this larger dataset have the potential to impose a more stringent constraint on the beyond Standard Model. From our work, we find that even with a median value of the J-factor, our results constraint the blue points in both mSUGRA and MSSM model and uncertainty band have already started to constraint the red points. Due to large uncertainty band in J factor of Tuc-II, maybe it is not possible to notice the significant improvement of constraints on DM models but our result gives a hint that by having a more detailed knowledge of internal structure, Tuc-II has a possibility to impose a very strong constraint on DM models in future. From our analysis results, we have shown that at above 100 GeV DM mass, the $<\sigma v>$ upper limit obtained from Tuc-II gives a more stringent limit than obtained from Ret-II. In future, with larger periods of data and with a more precise observation of the internal structure of Tuc-II, we should reduce its J-factor uncertainty band to a possible single upper limit curve of $<\sigma~v>$ and then Tuc-II might appear as one of the strongest DM dominated UFDs.\\

\begin{figure}[h!]

 { \includegraphics[width=0.6\linewidth]{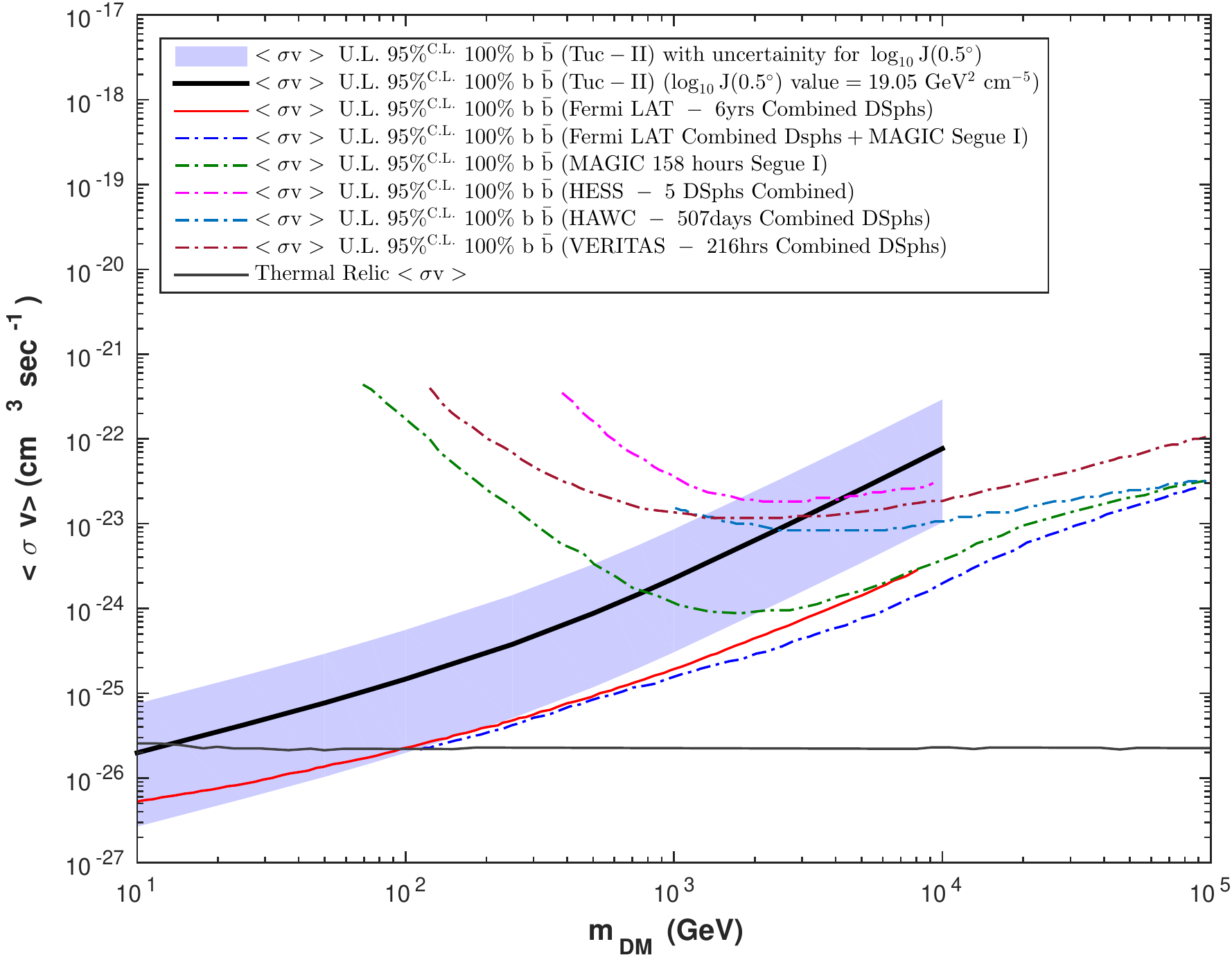}}
\caption{$<\sigma v>$ vs. $\rm{m_{DM}}$ for $b\bar{b}$ annihilation channel from Tuc-II (obtained in this paper) is shown in comparison with the results obtained from single or combined dSph studies by HESS~(\cite{bib:hess}), HAWC~(\cite{bib:hawc}), VERITAS~(\cite{bib:veritas}), MAGIC~(\cite{bib:fermag}), \textit{Fermi}-LAT~(\cite{bib:fermi}) and \textit{Fermi}-LAT+MAGIC~(\cite{bib:fermag}) collaboration, respectively. The shaded band represents the uncertainty in the DM density profiles in the Tuc-II. The horizontal dashed sky line corresponds to the relic abundance (or thermal) cross section estimated by Steigman et. al.~\cite{bib:steigman}.}
\label{fig:fig}
\end{figure}

\noindent A comparative study of the upper limits of WIMP pair-annihilation $<\sigma~v>$, as obtained in this paper, with the $<\sigma~v>$ obtained from other studies of the dSphs/UFDs is displayed in Fig.~11. It includes the results obtained from a combined analysis~\cite{bib:fermi} of 15 dSphs from six years of \textit{Fermi}-LAT data, the results from an analysis~\cite{bib:fermag} of $\gamma$-ray data from Segue-I, obtained earlier by using the Major Atmospheric Gamma-ray Imaging Cherenkov Telescopes (MAGIC) alone, as well as, the ones obtained from Segue-I by a joint \textit{Fermi}+MAGIC~\cite{bib:fermag} collaboration. Fig.~11 also includes the results from the analysis~\cite{bib:hess} of data from a combined study of $5$ dSphs by the High Energy Stereoscopic System (H.E.S.S.) of IACT telescopes and the results from the combined study of $15$ dSphs~\cite{bib:hawc} by the High Altitude Water Cherenkov (HAWC) gamma-ray observatory and of $4$ dSphs~\cite{bib:veritas} by the Very Energetic Radiation Imaging Telescope Array System (VERITAS), as well. In Fig.~11, we find that the joint \textit{Fermi}+MAGIC observations of Segue-I put the best overall limit on the DM annihilation $<\sigma~v>$ for a wide range of WIMP masses than the limits obtained by various other observations incorporated in that figure. As expected, the combined dSph analysis by the \textit{Fermi}-LAT collaboration also fairs well at masses up to about $1$~TeV, above which the \textit{Fermi}-LAT begins to suffer from low statistics. The comparison displayed in Fig.~11 seems to conform to the intuitive belief that the Cherenkov detector arrays should impose stringent limits on the annihilation $<\sigma~v>$ in the mass range $\gtrsim 10$~TeV, while a joint analysis of spaced-based + ground-based detector is likely to impose the most stringent limits on $<\sigma v>$ in the mass range $\lesssim 1$~TeV. It might be interesting to note that the upper-limits on $<\sigma~v>$ obtained by HAWC and by the \textit{Fermi}+MAGIC collaboration tend to converge in the mass range $\approx 100$~TeV, which might indicate that both the IACTs and the Water Cherenkov detectors are becoming competitive in regards to the DM search in the dSphs/UFDs. However, the limiting $<\sigma~v>$, displayed in Fig.~11, at $\lesssim 1$~TeV WIMP mass range is still about two orders of magnitude away from its relic abundance value. A coordinated effort to combine the data taken from several $\gamma$-ray telescopes, as well as, the enhancements of the sensitivity of the Cherenkov telescopes and the improvements of the data analysis techniques of the $\gamma$-ray telescopes in general seem, therefore, to be the pressing necessities.

                                                                                                                                                                                                                    \section*{A: T-TEST for unequal variance}
                                                                                                                                                                                                                   
\noindent T-TEST is a type of statistical hypothesis test which is used to determine whether there is any significant difference between the two groups \cite{bib:student, bib:ttest1, bib:ttest2}. The T-test is specially favored for smaller set of data (say, $n_{1}$ or/and $n_{2}$ $<$ 30)\cite{bib:student, bib:ttest1, bib:ttest2}. Under the null hypothesis, this test assumes that two sets of data come from the same or very likely population. The distribution of the T-TEST is symmetric, bell-shaped and very similar to the normal distribution. Hence, one of the key assumptions for T-test is that the variable of each sample is drawn from the normal distribution.\\
                                                                                                                                                                                                                    
                                                                                                                                                                                                                    \noindent For calculating the T-TEST, we need three values, such as i) mean, ii) standard deviation and iii) number of counts of each data set. The T-TEST produces two values as its output results, i) t-value and ii) degree of freedom (d.o.f.). The t-value is the same as test statistics (TS value). The test statistics is a generalized value which is evaluated from two data sets during the hypothesis test of T-TEST.\\
                                                                                                                                                                                                                    
                                                                                                                                                                                                                    \noindent There are three types of T-TESTs depending on the similarity and non-similarity of the standard deviations of two samples.
                                                                                                                                                                                                                    \noindent For our purpose (Fig.~3), we have compared the data from two independent samples. Our sample has unequal standard deviation and for this reason, we have used unequal variance T-TEST (also known as Welch's T-TEST)\cite{bib:welch, bib:welch1}. This T-TEST is mainly used when the two populations have different variances but the sample sizes of two data sets may or may not be equal. The t-value of unequal variance T-TEST is calculated as:  
                                                                                                                                                                                                                    
\begin{equation}
\rm{t-value} = \frac{mean_{1}-mean_{2}}{\sqrt{\frac{(var_{1})^{2}}{n_{1}}+\frac{(var_{2})^{2}}{n_{2}}}}
\end{equation}
                                                                                                                                                                                                               
\begin{equation}
                                                                                                                                                                                                              \rm{d.o.f.} = \frac{\Big(\frac{(var_{1})^{2}}{n_{1}}+\frac{(var_{2})^{2}}{n_{2}}\Big)^{2}}{\frac{(\frac{var_{1}^{2}}{n_{1}})^{2}}{n_{1}-1}+\frac{(\frac{var_{2}^{2}}{n_{2}})^{2}}{n_{2}-1}}
                                                                                                                                                                                                              \end{equation}
                                                                                                                                                                                                              
                                                                                                                                                                                                              \noindent Where,\\
                                                                                                                                                                                                              $mean_{1}$ and $mean_{2}$ = Average values of each of the sample sets\\
                                                                                                                                                                                                              $var_{1}$ and $var_{2}$ = Variance of each of the sample sets\\
                                                                                                                                                                                                              $n_{1}$ and $n_{2}$ = Number of records in each sample set \\ \\
                                                                                                                                                                                                              \noindent If we place our calculated t-value to the t-distribution (we have used the two-tailed t-distribution form), we can calculate the probability (p-value) associated with the t-value. The p-value allows us to evaluate whether the null-hypothesis is true. \\
                                                                                                                                                                                                              
\noindent The p-value associated with any distribution will vary between 0 to 1. For estimating the p-value, we need to assign a threshold value for p which is assigned as the significance level ($\alpha$) of the test. For our case, we have taken $\alpha$ = 5$\%$. The significance of the p-value can be interpreted in the following way:\\
                                                                                                                                                                                                             
                                                                                                                                                                                                              \noindent 1) The small p-value (i.e. $\preceq$ 0.05) indicates strong evidence against the null hypothesis, i.e. we can reject the argument of null-hypothesis.\\
                                                                                                                                                                                                              
                                                                                                                                                                                                              \noindent 2) The large p-value ($>$ 0.05) indicates weak evidence against the null hypothesis, i.e. we then fail to reject the null hypothesis.\\

                                                                                                                                                                                                          \noindent From Fig.~3(b), we always obtain the p-value $>$ 0.05 and it suggests that for positive bump and for the full energy range residual spectrum, we could not reject the null hypothesis. It also implies that we could not reject the possibility that dark matter model can provide a reasonable fit to the residual energy spectrum, even for full energy range. \\
                                                                                                                                                                                                              
                                                                                                                                                                                                              \section*{B: Normality test of dataset}
                                                                                                                                                                                                              \noindent One of the key assumptions for two sample T-TEST is that both the samples should follow the normal distribution \cite{bib:student, bib:ttest1, bib:ttest2}. There are several statistical tests to check the normality of data, such as the Shapiro–Wilk \cite{bib:shapiro} or Kolmogorov–Smirnov test \cite{bib:kol, bib:smir} or the normal quantile plot \cite{bib:qq1}. We have used the quantile-quantile (Q-Q) plot to check the normality of our dataset.\\
                                                                                                                                                                                                              
                                                                                                                                                                                                              \noindent Quantile-quantile (Q-Q) plot is a graphical representation which helps us to determine whether two data sets originate from the populations with a common distribution \cite{bib:qq1, bib:qq2}. This plot displays the quantiles of the sample data versus the theoretical quantile values from a normal distribution \cite{bib:qq1, bib:qq2}. \\ 
                                                                                                                                                                                                                                                                                                                                                                                                                            
                                                                                                                                                                                                             \noindent In normal Q-Q plot, the quantiles from a theoretical normal distribution are plotted on the horizontal axis and the corresponding quantiles from the experimental data are plotted on the y-axis. For normaly distributed sample data, the points plotted in the Q-Q plot should fall approximately on a straight line indicating the high positive correlation between the quantiles of the theoretical normal data and sample data \cite{bib:qq1, bib:qq2}.\\
                                                                                                                                                                                                                                                                                                                                                                                                                                                                                                                                                                                                                                     
                                                                                                                                                                                                                                                                                                                                                                                                                                                                                                                                                                                                                                     \noindent For testing the correlation between theoretical normal data and sample data, the paired data in Q-Q plot is generally fitted to a straight line and that fitting returns the regression equation (y=ax) along with the coefficient of determination as $R^{2}$. Pearson correlation coefficient (r) is computed from $R^{2}$, such as r=$(R^{2})^{1/2}$=R \cite{bib:rr1, bib:rr2}. Theoretically, r-values between 0.9 to 1 indicate a high correlation between two data sets \cite{bib:rr1, bib:rr2}. The lesser the deviation from normality, the closer the r-value will be to 1 and the r-value=0 shows there is no association between two samples \cite{bib:rr1, bib:rr2}.\\
                                                                                                                                                                                                                                                                                                                                                                                                                       
                                                                                                                                                                                                                                                                                                                                                                                                                        \noindent No data set indeed follow the exact normal distribution but most of the statistical analysis requires an approximately normally distributed population. For our analysis, we have checked the Q-Q plot and the corresponding r-value for all the samples that we have used for T-TEST. The Q-Q plot of our samples, i.e. i) for full energy range residual spectrum and ii) for positive bump in residual spectrum, both follow the straight line and all of them provide the r-value $>$ 0.94. We can then state that our data sets are not free from deviation but such high r-value indicates that our sample almost follows the normal distribution. Hence, we can appropriately apply the T-TEST goodness of fitting to our dataset.\\

\section*{Acknowledgement}
\noindent We are thankful to the anonymous referee for mentioning some important points which helped us to improve our paper a lot. PB is grateful to DST INSPIRE Fellowship Scheme for providing the support for her research. PB would like to thank Dr. Chowdhury Aminul Islam and Prateek Chawla for their valuable discussion. 
We would like to thank Dr. Stefano Profumo and Dr. Tesla Jeltema of University of California, Santa Cruz, U.S.A., for providing useful guidance in \textit{Fermi}-LAT data analysis. We are particularly grateful to the \textit{Fermi} Science Tools for allowing us to freely access the \textit{Fermi}-LAT data and providing all the needed guidance for our analysis.

\end{document}